\documentclass[journal]{IEEEtran}
\usepackage[utf8]{inputenc}
\usepackage{graphicx}
\usepackage{amsmath}
\usepackage{amsthm}
\usepackage{amssymb}
\usepackage{import}
\usepackage{color}
\usepackage{array}
\usepackage{diagbox}
\usepackage{subfig}
\usepackage{multicol}
\usepackage{multirow}
\usepackage{placeins}
\usepackage[linesnumbered,ruled]{algorithm2e}
\usepackage{hyperref}
\usepackage{balance}
\usepackage[normalem]{ ulem }
\usepackage{soul}
\title{Estimation of urban traffic state with probe vehicles}

\author{
\IEEEauthorblockN{Cyril Nguyen Van Phu\IEEEauthorrefmark{1}, Nadir Farhi} \\
\IEEEauthorblockA{COSYS-GRETTIA, Univ Gustave Eiffel, IFSTTAR, F-77454 Marne-la-Vallée, France}\\
\IEEEauthorblockA{\IEEEauthorrefmark{1} corresponding author}
}

\newtheorem{prop}{Proposition}
\newtheorem{defn}{Definition}
\begin{document}
\maketitle
\begin{abstract}
We present in this paper a method to estimate urban traffic state with communicating vehicles.
Vehicles moving on the links of the urban road network form queues at the traffic lights. 
We assume that a proportion of vehicles are equipped with localization and communication capabilities, and name them probe vehicles.
First, we propose a method for the estimation of the penetration ratio of probe vehicles, as well as the vehicles arrival rate on a link.
Moreover, we show that turn ratios at each junction can be estimated.
Second, assuming that the turn ratios at each junction are given,
we propose an estimation of the queue lengths on a 2-lanes link, by extending a 1-lane existing method.
Our extension introduces vehicles assignment onto the lanes.
Third, based on this approach, we propose control laws
for the traffic light and for the assignment of the arriving vehicles onto the lane queues.
Finally, numerical simulations are conducted with Veins framework that bi-directionally couples microscopic road traffic and communication simulators.
We illustrate and discuss our propositions with the simulation results.
\end{abstract}

\begin{IEEEkeywords}
Intelligent transportation systems, Queuing systems.
\end{IEEEkeywords}

\section{Introduction}
\label{sec:intro}
\subsection{State of the art}
Different techniques are traditionally used to measure road traffic parameters; for example we can cite inductive loops or video cameras.
There is nowadays an infrastructure-less technique to estimate traffic flow parameters such as queue lengths : GPS localization system coupled with communicating vehicles, namely probe vehicles.
This kind of equipment penetration ratio is increasing and does not need heavy set up.

Probe vehicles were historically studied for measuring travel times \cite{doi:10.1080/15472450701849667}.
They also helped to estimate penetration ratio and arrival rate of vehicles (equipped and non equipped) on a link.
For example, the author of~\cite{COMERT2016502} derived these estimations from the estimation of queue lengths at junctions, queue lengths being estimated using the information provided by the probe vehicles.
Thus, we can see that in order to characterize urban road traffic state and its primary parameters such as arrival rates or penetration ratio, estimating queue lengths at junctions is an important step.
Furthermore, Varaiya~\cite{VARAIYA2013177} has modeled a road network as ''a controlled store-and-forward (SF) queuing network`` 
and proposed an algorithm to control this network of queues.
Indeed, minimizing delays and waiting times can be done by minimizing queue lengths at junctions controlled with traffic light signals.
Hence, queue length estimation is a major measurement input data, used to control traffic light signals, and so transportation road networks.

Concerning queue lengths, in 1963, Miller~\cite{miller1963settings} found an approximation of the average queue length at junctions.
More recently, the authors of~\cite{6728492} used shockwave theory to refine queue length estimation.
Some works also proposed to use probability distribution of the queues \cite{HEIDEMANN1994377}.
Other works used Markov chains to model the dynamics of queue lengths \cite{viti2004modeling}.
The authors of~\cite{COMERT2009196}~and~\cite{ROSTAMISHAHRBABAKI2018525} have addressed the queue length estimation with probe vehicles by proposing a probabilistic analytical model.
In~\cite{COMERT2009196}, the authors have estimated queue length in under-saturated traffic conditions,
with the ``a priori knowledge of the marginal distribution of the queue length'' and using
``the location information of the last probe vehicle in the queue``.
The authors of~\cite{ROSTAMISHAHRBABAKI2018525} have proposed a method to estimate the queue length, the incoming arrival rate, and the output flow, on a m-lanes link ($m\geq 2$).
The estimations are given for low or saturated demand with no requirement of information concerning the timings of the traffic light signal.
In~\cite{ROSTAMISHAHRBABAKI2018525}, all the lanes are assumed to be balanced (i.e. cars share the lanes of the link without any preference).
Therefore, all the lanes would have the same length.
In~\cite{ZHENG2017347}, the authors estimate arrival rate for low penetration ratio of equipped vehicles.
The method proposed in~\cite{ZHENG2017347} uses as input data ``vehicle trajectories approaching to an intersection as well as traffic signal status''.
The trajectories of equipped vehicles are used to detect if a probe vehicle has stopped at the traffic light and its stopping position.
With these information,  the arrival rate is estimated and bounds for this arrival rate are given.
In~\cite{Tiaprasert7053921}, the authors proposed another method. They have lower-bounded the queue length by
``the location of the last stopped connected vehicle'' and upper-bounded it, when the bound exists, by the location of the 
``closest  moving connected vehicle''. Once bounded, the queue length is estimated using the least-mean-square-error method and 
the noise is filtered using discrete wavelet transform.
In 2015, the authors of~\cite{LEE20151} have addressed the two lanes case by combining discriminant models ``based on time occupancy rates and impulse memories'' from detectors.
The proportions of total traffic volume in each lane are estimated with Kalman filter.
In 2018, the authors of~\cite{doi:10.1080/15472450.2017.1300887} have also addressed the two lanes case.
They have measured `` individual probe vehicles’ shockwave speed''.
Then the lane each probe is moving on, is determined by discriminating the two lanes with data clustering methods.
They have shown that a bivariate mixture model clustering gives the best results.
Shockwave theory and LWR (Lighthill, Whitham and Richards) model~\cite{Lighthill317, doi:10.1287/opre.4.1.42} refine the queue length estimation.
\subsection{Paper contribution and organization}
We present here an extension of an existing method that uses probe vehicles for the estimation of urban traffic state, including
penetration ratio of communicating vehicles, vehicular arrival rates, as well as the queue lengths of an urban link.
Our extension considers the general case where different destinations can be associated to the lanes, which produces different arrival rates
to each lane of the urban link. We propose here to estimate the joint probability distribution of all the queue lengths of the urban link,
instead of estimating only one queue length for the link, as done in~\cite{ROSTAMISHAHRBABAKI2018525,ZHENG2017347}.
This distinction of the lane queues improves the estimation of the number of cars on the queues. Moreover, it gives us the possibility
to control the flows of each queue separately, and then ameliorates the traffic control on the junction.
We propose in addition control laws for balancing the queue lengths in a multi-lane link.
We present here our method on a link of two lanes.
We think that the ideas presented in this particular case could be adapted in order to address the general case ($m$-lanes, with $m>2$).
Furthermore, we think the method proposed here could be used and extended in a decentralized manner to the network case because of the low computational 
effort needed for the one link case.
The estimators and the control laws we propose here would permit us to perform multi-level urban traffic control,
as initiated in~\cite{NguyenVanPhu8005594} (local control) and in~\cite{FARHI201541} (semi-decentralized control).
 
In section~\ref{sec:intro} we give an introduction with the related works. 
In section~\ref{sec:problem} we describe the problem statement and the notations.
In section~\ref{sec-traffic-state} we propose an estimation of traffic state parameters : penetration ratio of probe vehicles, vehicles arrival rate
(subsection~\ref{sub-primary}), and queue lengths in the case of two incoming lanes (subsection~\ref{sub-queue_2_lanes}).
Our method estimates queue lengths at junctions with two lanes incoming roads, under the hypothesis of under-saturated traffic (moderate/low demand without overflow queue).
We also assume that the GPS localization system is not able to determine which lane a vehicle is moving on because of a typically five meters accuracy~\cite{van2015world}.
We extend the analytic model proposed in~\cite{COMERT2009196} to the two-lanes case by introducing a vehicle assignment model onto the lanes.
In subsection~\ref{sub-assign}, we propose a control of the traffic light and an optimal assignment of the vehicles onto the lanes, in order to balance the two lanes queue lengths.
In section~\ref{sec-results}, numerical simulations are conducted with Veins framework \cite{sommer2011bidirectionally}
which bi-directionally couples microscopic road traffic and communication simulators.
Finally, we conclude in section~\ref{sec-conclusion}.

\section{Problem statement}
\label{sec:problem}
In this section we describe the main assumptions of our work and the notations used.
\subsection{Assumptions}
\begin{figure}[htbp]
  \begin{center}
    \includegraphics[width=0.7\linewidth, keepaspectratio]{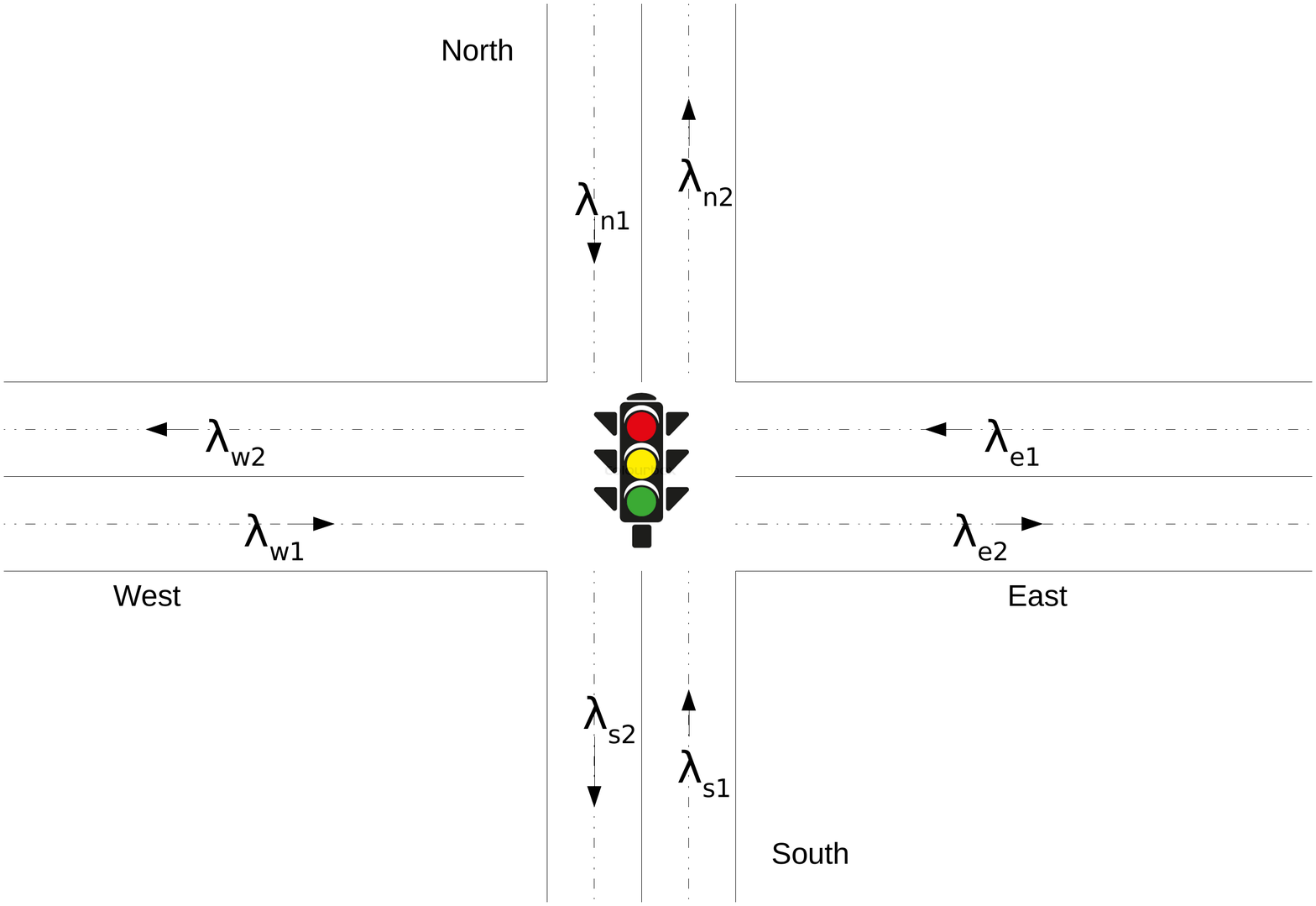}
    \caption{A signalized junction of the road network.}
    \label{fig:junction}
  \end{center}
\end{figure}

\paragraph{Road network topology}
We consider a road network composed of junctions controlled by traffic light signals, and links between junctions.
We assume that all the incoming and outgoing links to/from a signalized junction have maximum two lanes.
We define an entry link of the network as a link which does not have a start node.
We assume that the geometry of the road network is known.
A typical junction is represented on Fig.~\ref{fig:junction}.
 We also assume that the timings of the traffic light signal are known, and specially the red times.
\paragraph{Traffic demand}
We assume that the travel demand is exogenous, which means that the demand is located only at the entry links of the network.
We assume that the vehicles arrive onto each link $l$ under a Poisson process of rate $\lambda_l$.
We consider in this paper the low/moderate demand case where the Poisson arrival assumption is valid.
In~\cite{COMERT2016502} the author discusses the Poisson arrivals assumption and recalls that this assumption is commonly used to describe arrivals at isolated intersections,
specially in the case of low/moderate demand with no overflow queue.
The vehicles form queues at junctions.
Since we assume Poisson arrivals, we consider that the queues are empty at the beginning of each red time (no overflow queue).

\paragraph{Probe vehicles}
We assume that a ratio $p$ (with $0\leq p \leq1$) of vehicles are equipped with localization and communication systems and we name them probe vehicles.
The probe vehicles send their positions and speeds to a road side unit (RSU) coupled with the traffic light signal of the junction.
We assume that the transmit power of the communication system embedded in every vehicle is strong enough, and that the sensitivity of the RSU is accurate enough,
such that the RSU can detect every vehicle in every incoming or outgoing link of its associated junction.
We consider the case where the localization system embedded in the vehicles is not accurate enough to discriminate the lane the vehicle is moving on.

\begin{figure}[htbp]
  \begin{center}  
    \includegraphics[width=0.9\linewidth, keepaspectratio]{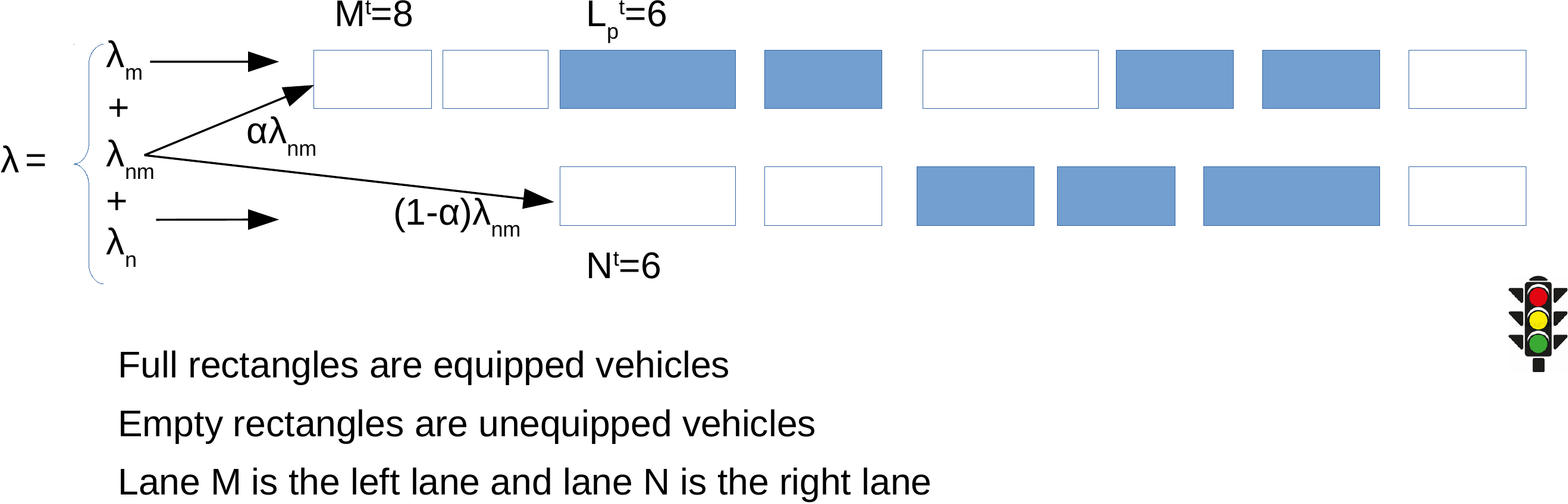}  
    \caption{Queues in 2-lanes incoming link. Vehicles that can choose both lanes are assigned onto lane $M$ with probability $\alpha$ and onto lane $N$ with probability $(1-\alpha)$.}
    \label{fig:queue_2_lanes}
  \end{center}
\end{figure}

\paragraph{Turn ratios}
Fig.~\ref{fig:queue_2_lanes} represents the queues we consider, on a link of the road network.
Probe vehicles are represented by full rectangles and unequipped vehicles are represented by empty rectangles.
Some vehicles are necessarily assigned to the queue on lane $N$ (they turn right) and other vehicles are necessarily assigned to the queue on lane $M$ (they turn left).
Vehicles going straight can choose both lanes. We will assume that a ratio $\alpha$ (with $0\leq\alpha\leq1$) of such vehicles going straight will choose
the queue on lane $M$.
So, on a 2-lanes link, we assume that the main flow $\lambda$ is composed of three flows :  
\begin{enumerate}
 \item the flow with arrival rate $\lambda_n$ which is necessarily assigned to lane $N$ (vehicles turning right).
 \item the flow with arrival rate $\lambda_m$ which is necessarily assigned to lane $M$ (vehicles turning left).
 \item the flow with arrival rate $\lambda_{nm}$ which can be assigned to both lanes $N$ or $M$ (vehicles going straight).
\end{enumerate}
We consider that these three flows are independent and identically distributed (iid) stochastic arrivals, each one being a Poisson process.
We suppose that a fraction $\alpha$ (with $0\leq\alpha\leq1$) of the flow $\lambda_{nm}$ is assigned to lane $M$ and 
the complement $(1-\alpha)$ of this same flow $\lambda_{nm}$ is assigned to lane $N$.
As denoted in TABLE~\ref{tab:notations}, $A^t$ is the random variable representing the assignment onto the two lanes.
We assume that $A^t$ is following a Bernoulli law such that $P(A^t=1)=\alpha$ and $P(A^t=0)=1-\alpha$.
Thus, $\mathbb E(A^t)=\alpha$.
We define : 
\begin{align}
  & \mu_N(t):=r_N(t)(\lambda_n+(1-\alpha)\lambda_{nm}) \\
  & \mu_M(t):=r_M(t)(\lambda_m+\alpha\lambda_{nm})
\end{align}

We will show later in Proposition~\ref{prop1} that $\mu_N(t)$ and $\mu_M(t)$ represent the average arrival rate multiplied by the red duration on respectively lane $N$ and lane $M$.

Also, we assume that the turn ratios are given. Indeed, it is easy to measure the turn ratios as following :
the RSU detects all the probe vehicles in its radio range area. This is because the probe vehicles embed WAVE (Wireless Access in Vehicular Environments)~\cite{Wave} on OBU (on board unit).
In~\cite{Wave}, the basic safety messages (BSM) broadcast periodically the location and speed of probe vehicles.
This is a default feature which is also implemented as a basic function in VEINS simulator~\cite{sommer2011bidirectionally} that we use in this paper.
So, if we know at time $t$ the location of each probe vehicle on a given link and its unique identifier, it is enough to look at a time $t+t_x$ ($t_x$ being a time shift), where those vehicles are located.
With this method, it is possible to estimate the turn ratios $l_n$, $l_m$ and $l_{nm}$ which are the proportions of the main flow $\lambda$ on the incoming link that respectively turn right, left or go straight.
We will note : 
$\lambda_n = l_n \lambda$,
$\lambda_m = l_m \lambda$,
$\lambda_{nm} = l_{nm} \lambda$,
with $l_n+l_m+l_{nm}=1$.
We assume in this paper that $l_n$, $l_m$, and $l_{nm}$ are given.
\subsection{Notations}
We will use the notations of TABLE~\ref{tab:notations}.
\begin{table}[htbp]
  \begin{tabular}{|l|p{6.4cm}|}
    \hline
    Name & Definition \\
    \hline
    $L_V$ & the average vehicle length \\
    $G_V$ & the minimum distance gap between vehicles\\
    $R$ & the total red time in one cycle \\
    $r_N(t)$ & the time since the beginning of the red phase for lane N (it is $0$ if we are not in red phase at time $t$), $0\leq r_N(t)\leq R$. \\
    $r_M(t)$ & the time since the beginning of the red phase for lane M (it is $0$ if we are not in red phase at time $t$), $0\leq r_M(t)\leq R$. \\
    $\lambda_n,\lambda_m$,$\lambda_{nm}$ & the average arrival rate in vehicles/second for vehicles that respectively turn right, left or go straight.\\
    $\lambda=\lambda_n+\lambda_m+\lambda_{nm}$ & the total arrival rate for the incoming link in vehicles/second.\\
    $l_n, l_m, l_{nm}$ & the proportions (turning ratios) of the main flow $\lambda$ on the incoming link that respectively turn right, left or go straight, with
        $\lambda_n = l_n \lambda$, $\lambda_m = l_m \lambda$, $\lambda_{nm} = l_{nm} \lambda$\\
    $x(t)$ & the total number of vehicles on all the lanes of the considered link at time $t$. \\
    $x_{p}(t)$ & the number of probe vehicles on all the lanes of the considered link at time $t$. \\
    $p$, $0 \leq p \leq 1$ & the penetration ratio of probe vehicles. \\
    $N^t$ & the total number of vehicles in the queue at time $t$ and lane N. In this paper, $N^t$ is assumed to be a random variable. \\
    $M^t$ & the total number of vehicles in the queue at time $t$ and lane M. In this paper, $M^t$ is assumed to be a random variable. \\
    $A^t$ & the assignment of a vehicle entering the edge at time $t$. $A^t=1$ if the vehicle is assigned on lane M and $A^t=0$ if it is assigned on lane N. $A^t$ is assumed to be a random variable.\\
    $L_p^t$ & the location (in number of vehicles) of the last probe in the queue, namely the last connected vehicle, at time $t$. $L_p^t$ is assumed to be a random variable, taking value $l_p$.\\
    $N_p^t$ & the total number of probe vehicles in the queue at time $t$ and lane N.\\
    $M_p^t$ & the total number of probe vehicles in the queue at time $t$ and lane M.\\
    $c_p$ & the total number of probe vehicles in all the lanes and all the queues at time $t$.\\
    \hline
  \end{tabular}
  \caption{Notations}
  \label{tab:notations}
\end{table}

\section{Traffic state estimation}
\label{sec-traffic-state}
\subsection{Primary parameters estimation}
\label{sub-primary}

In this section, we give a method for the estimation of the primary traffic parameters $p$ and $\lambda$.
We assume that every probe vehicle in the RSU radio range area is assigned to an incoming or outgoing link to/from the junction. 
Thus, the total number of incoming probe vehicles $x_p(t)$ in a given link to the junction is known.

We consider vehicles $i$ moving at speed $v_i(t)$ and at a distance $\rho_i(t)$ (depending on time $t$) from the traffic light.
Let us consider the following definition.
\begin{defn}
For a given threshold car-speed $v^*$ and a given threshold car-distance $\rho^*$ to the junction,
the vehicles queue $Q=Q(t,v^*,\rho^*)$ is defined by $Q=\{i, v_i<v^* \text{ and } \rho_i<\rho^* \}$.
\end{defn}
$\rho^*$ is useful because if the queue would exceed the bound $\rho^*$, we could know that the assumption of low/moderate demand is not adequate. 
Furthermore, $\rho^*$ is less than the edge length, so the queue keeps bounded.
We then denote by $Q_p$ the subset of $Q$ that includes only probe vehicles, $Q_p\subset Q$.
The total number of probes $c_p$ in the queue is given by the cardinal (number of elements) of the set $Q_p$.
We assume mixed vehicles (equipped and non equipped) with an average vehicle length $L_V$ and minimum distance gap $G_V$ between vehicles.
$\rho_0$ denotes the offset distance from the RSU to the stop line of the traffic light signal.
We propose to compute $l_p$ the last probe location in the unit of ''number of vehicles`` as follows. We have :

\begin{equation}
\max_{i \in Q_p}(\rho_i)=\rho_0+ l_pL_V+(l_p-1)G_V
\end{equation}
Then,
\begin{equation}
l_p= [(\max_{i \in Q_p}(\rho_i)-\rho_0 +G_V)/(L_V+G_V)]
\end{equation}
where $[\cdot]$ denotes the round operator to the nearest integer.

Given $c_p$, $l_p$, Comert~\cite{COMERT2016502} has derived many estimators for $p$, one of them being $c_p/l_p$, which is biased for $p<1$.
We follow here the same idea and propose a variation of the estimator of $p$.
For the one lane case, we propose:
\begin{equation}
\hat{p}= (c_p-1)/(l_p-1), \text{ for } l_p>1
\label{eq:p}
\end{equation}
We have : $N^t = l_p + 1/p - 1$, where $1/p-1$ represents the average backlog of the queue behind the last probe. 
Then $\hat{p} = c_p/N^t = c_p/(l_p + 1/p -1)$. Moreover, by following the same arguments of the proof in~\cite{COMERT2016502}, 
it is easy to check that this estimator is unbiased for every $p$, i.e. $\mathbb E(\hat{p}) = p, \forall p, 0\leq p\leq 1$.

For two lanes, we introduce : \\
$\kappa:=\min(\mu_n,\mu_m)/\max(\mu_n,\mu_m)$.
We can see that $\kappa$ does depend only on turn ratios but not on the arrival rate for the link, because $\kappa$ is a ratio.
 \begin{equation}
  \kappa=\frac{\min(r_N(t)(l_n+(1-\alpha)l_{nm}),r_M(t)(l_m+\alpha l_{nm}))}{\max(r_N(t)(l_n+(1-\alpha)l_{nm}),r_M(t)(l_m+\alpha l_{nm}))}
 \end{equation}
We consider queue lengths on lanes $N$ and $M$ respectively equal to $n$ and $m$.
We propose :
 \begin{equation}
 \hat{p}=\frac{c_p}{n+m}
  \label{eq:p2}
  \end{equation}
  By the way, in our case, the length $n$ of queue $N$ can be estimated with the number of arrivals on lane $N$ during $r_N(t)$ which is $\mu_n$.
 As $\mu_n+\mu_m=\max(\mu_n,\mu_m)+\min(\mu_n,\mu_m)$ and by estimating $\max(\mu_n,\mu_m)=l_p+1/p-1$, 
 where $1/p-1$ represents the backlog of the queue behind the last probe,  
 we can write :
 \begin{equation}
 \mu_n+\mu_m=\max(\mu_n,\mu_m)\left(1+\frac{\min(\mu_n,\mu_m)}{\max(\mu_n,\mu_m)}\right)
  \end{equation}
   \begin{equation}
 \mu_n+\mu_m=(l_p+1/p-1)(1+\kappa)
   \label{eq:mu}
  \end{equation}
  We introduce $c_\kappa=c_p/(1+\kappa)$ and replace~(\ref{eq:mu}) in~(\ref{eq:p2}).
  Finally, we get the following equation : 
  \begin{equation}
    \hat{p}=\frac{c_\kappa}{l_p+(1-p)/p}
  \end{equation}
  Hence, solving in $\hat{p}$, by putting $p=\hat{p}$ :
      \begin{equation}
          \hat{p}=(c_\kappa-1)/(l_p-1), \text{ for } l_p>1 \text{ and } c_p>1
          \label{eq:hatp2}
      \end{equation} 
  which extends~(\ref{eq:p}) for the case of two lanes.
  Similarly, we can check that this estimator for two lanes is unbiased, by following again the same arguments as in~\cite{COMERT2016502}.
  In Appendix~\ref{Appendix:p}, we propose another method to compute $\hat{p}$ which is based on the calculus of an unbiased estimator
  and gives the same result of formula~(\ref{eq:hatp2}).

We propose to compute $\lambda$ with formula~(\ref{eq:lambda2}) by simply accumulating probe vehicles on the entire radio range area of the RSU during red time,
and using $\hat{x}=x_p/p$.
$\lambda$ should be computed when all the lanes of the considered link have a red light at the traffic light.
We formulate this as following, 
where $t_0$ is the starting time for the red light on both lanes and $t_1$ is the time after which the set of the two lanes are not at red light.
\begin{equation}
  \hat{\lambda} = \frac{x_p(t_1)-x_p(t_0)}{p(t_1-t_0)}
  \label{eq:lambda2}
\end{equation}

\subsection{Queue length estimation}
\label{sub-queue_2_lanes}
Once $p$ and $\lambda$ are estimated with probe vehicles, we can refine our traffic state estimation (queue lengths).
We propose in this section to estimate all the queue lengths associated to all the lanes on a link of the road network.
We will propose a model that uses vehicular assignment onto the lanes, for links composed of two incoming lanes.
In a first step, an analytical probability distribution formulation of the queue lengths, without using the information from the probe vehicles, will be presented.
Then, we will use the information provided by the probe vehicles :
while generalizing the work for 1-lane road done in~\cite{COMERT2009196} to the 2-lanes case, we will refine our analytical formulation.
We recall here that we can not directly detect the lanes on which the probe vehicles are moving because of insufficient accuracy of GPS localization system~\cite{van2015world}, which makes the problem not obvious.
\paragraph{Distribution probability law of the 2-lanes without having the information provided by the probes}
\label{sub-marginal}
We first propose an estimation of the probability distribution $P(N^t=n,M^t=M)$ without having any information from the probe vehicles.

\begin{prop}
  \label{prop1}
  \begin{multline}
    P(N^t=n,M^t=m)=\frac{\mu_N(t)^{n}e^{-\mu_N(t)}}{n!}\frac{\mu_M(t)^{m}e^{-\mu_M(t)}}{m!} \\    
  \end{multline}
\end{prop}
\proof
We subdivide the Poisson process of rate $\lambda_{nm}$ common to the two lanes.
The common arrival of rate $\lambda_{nm}$ is splitted with probability $\alpha$ to lane $M$ and probability $(1-\alpha)$ to lane $N$.
The two produced flows are independent random flows each one following Poisson process of parameters respectively
$\alpha\lambda_{nm}$ for the flow assigned to lane $M$ and 
$(1-\alpha)\lambda_{nm}$ for the flow assigned to lane $N$.
Furthermore, the splitted Poisson processes are independent;
see subdividing Poisson process in reference~\cite{gallager2013stochastic}.

By combination, arrivals on lane $N$ is the sum of two independent Poisson processes. 
Using the stationary property of Poisson processes, we can show that the number of arrivals in $[0,r_N(t)]$ on lane $N$ is a Poisson process of parameter $\mu_N$.
Similarly, the number of arrivals on lane $M$ in $[0,r_M(t)]$ is a Poisson process of parameter $\mu_M$.
As these two arrival flows on lanes $N$ and $M$ are independent, 
then the bivariate distribution probability law of the two queue lengths is the product of two Poisson Law of parameters $\mu_N$ and $\mu_M$.
\endproof

\paragraph{Distribution probability law of the 2-lanes queue lengths with the information provided by the probe vehicles}
\label{sub-proba}
We present here the conditional probability law of the two queue lengths, taking into consideration the information provided by the probe vehicles, 
specially the location of the last probe $l_p$ and the total number of probes in the two lanes queues $c_p$.
We recall here $N_p^t$ and $M_p^t$ are the number of probe vehicles respectively on the lane $N$ and on the lane $M$, at time $t$.
\vspace{0.85cm}
\begin{prop}
\label{prop-2}
\hspace{2em}
  \begin{itemize}
    \item If $lp \leq\max(n,m)$ and $c_p \leq n+m$, then
      \begin{multline}
	P(N^t=n,M^t=m|L_p^t=l_p,N_p^t+M_p^t=c_p)=\\
	\frac{\binom{l_p-1+\min(l_p,n,m)}{c_p-1}(1-p)^{n+m}P(N^t=n,M^t=m)}{\sum \limits_{\substack{j,k\geq0 \\ 
	               \text{subject to} \\ \max(j,k)\geq l_p\\ j+k \geq c_p\\}}^{}\binom{l_p-1+\min(l_p,j,k)}{c_p-1}(1-p)^{j+k} P(N^t=j,M^t=k)}. \nonumber    
      \end{multline}
    \item Otherwise, \\ $P(N^t=n,M^t=m|L_p^t=l_p,N_p^t+M_p^t=c_p)=0$.
  \end{itemize}
\end{prop}

\proof By Bayes' rule we have
\begin{multline}
  P(N^t=n,M^t=m|L_p^t=l_p,N_p^t+M_p^t=c_p)=\\
  \frac{P(N^t=n,M^t=m,L_p^t=l_p,N_p^t+M_p^t=c_p)}{P(L_p^t=l_p,N_p^t+M_p^t=c_p)}
  \label{eq:2lanesA}
\end{multline}

Then the numerator in~(\ref{eq:2lanesA}) is written
\begin{multline}
  P(N^t=n,M^t=m,L_p^t=l_p,N_p^t+M_p^t=c_p)=\\
  P(L_p^t=l_p|N_p^t+M_p^t=c_p,N^t=n,M^t=m)\\
  P(N_p^t+M_p^t=c_p|N^t=n,M^t=m)\\
  P(N^t=n,M^t=m)
  \label{eq:2lanesB}
\end{multline}

We have
\begin{itemize}
 \item $P(L_p^t=l_p|N_p^t+M_p^t=c_p,N^t=n,M^t=m) = \\~~\\ \binom{l_p-1+\min(l_p,n,m)}{c_p-1} / \binom{n+m}{c_p}$. \\
 \item $P(N_p^t+M_p^t=c_p|N^t=n,M^t=m) = \\~~\\ \binom{n+m}{c_p}p^{c_p}(1-p)^{n+m-c_p}$.
\end{itemize}

For the calculus of
$P(L_p^t=l_p|N_p^t+M_p^t=c_p,N^t=n,M^t=m)$, we followed the same ideas as those of section~3 in~\cite{COMERT2009196}.
Indeed, we will use the example of Fig.~\ref{fig:queue_2_lanes} where $l_p=6$, $M^t=8$, $N^t=6$ and $c_p=7$.
The probability is then computed by selecting the total number of events where $L_p=l_p=6$ divided by the sample space.
The sample space, which is composed of all the last probe possible locations is given by $\binom{n+m}{c_p}=\binom{14}{7}$.
For $L_p=l_p=6$ we must have all the probes in the preceding locations.
The event space has a number of events corresponding to choosing $(c_p-1)=6$ probes among $l_p-1+\min(l_p,n,m)=6-1+6=11$ positions available.
Here, the event space has a total number of elements given by $\binom{l_p-1+\min(l_p,n,m)}{c_p-1}=\binom{11}{6}$.
This is why : 
\begin{multline}
P(L_p^t=l_p|N_p^t+M_p^t=c_p,N^t=n,M^t=m) = \\
\binom{l_p-1+\min(l_p,n,m)}{c_p-1} / \binom{n+m}{c_p}
\end{multline}

For the calculus of $P(N_p^t+M_p^t=c_p|N^t=n,M^t=m)$ we have $c_p$ probe vehicles among $n+m$ total vehicles.
The probability for a vehicle to be a probe vehicle is $p$ and the probability to be unequipped is $(1-p)$.
The configurations considered in this case are $c_p$ vehicles equipped and $(n+m-c_p)$ vehicles unequipped.
The number of combinations of such configurations is $\binom{n+m}{c_p}$.
This is why : 
\begin{multline}
P(N_p^t+M_p^t=c_p|N^t=n,M^t=m) = \\ 
\binom{n+m}{c_p}p^{c_p}(1-p)^{n+m-c_p}
\end{multline}
So the numerator in~(\ref{eq:2lanesA}) is given by : 
\begin{multline}
  P(N^t=n,M^t=m,L_p^t=l_p,N_p^t+M_p^t=c_p)=\\
  \binom{l_p-1+\min(l_p,n,m)}{c_p-1} 
  p^{c_p}(1-p)^{n+m-c_p}\\
  P(N^t=n,M^t=m) \nonumber
\end{multline}
The denominator in~(\ref{eq:2lanesA}) is the marginal distribution probability of $P(N^t=j,M^t=k,L_p=l_p,N_p^t+M_p^t=c_p)$ on $(j,k)$.
Therefore, the ideas to compute this probability are the same as the ideas used to compute the numerator of~(\ref{eq:2lanesA}).
We notice here that the last probe position (in the unit number of vehicles) is necessarily less than or equal to the maximum of the queue lengths,
since the last probe is necessarily in one of the two lanes queues.
Similarly, the total number of probes $c_p$ is less than or equal to the total number of vehicles in the queues,
since the probes are in the queues.
Therefore, we can write :
\begin{multline}
  P(L_p^t=l_p,N_p^t+M_p^t=c_p)=\\
  \sum \limits_{\substack{j,k\geq0 \\ \max(j,k)\geq l_p\\ j+k \geq c_p\\}} P(N^t=j,M^t=k,L_p=l_p,N_p^t+M_p^t=c_p). \nonumber
\end{multline}
\endproof
\paragraph{Estimators}
\label{sub-est}
The distribution probability law of the couple $(N^t,M^t)$ is known;
see Proposition~\ref{prop-2} .
As, $\mathbb E(N^t,M^t)=(\mathbb E(N^t),\mathbb E(M^t))$, one way to estimate the two queue lengths is to derive each queue length separately from the couple, by computing the expectation of $N^t$ and $M^t$ separately.
We propose the following estimator for queue length on lane $N$ :
\begin{multline}
\label{eq:estimator}
\mathbb E(N^t|L_p^t=l_p,N_p^t+M_p^t=c_p)=\\
\sum_{n\geq0}n\sum_{k\geq0}^{}P(N^t=n,M^t=k|L_p^t=l_p,N_p^t+M_p^t=c_p)
\end{multline}
Similarly for the queue length on lane $M$, we will choose :
\begin{multline}
\label{eq:estimator2}
\mathbb E(M^t|L_p^t=l_p,N_p^t+M_p^t=c_p)=\\
\sum_{m\geq0}m\sum_{j\geq0}^{}P(N^t=j,M^t=m|L_p^t=l_p,N_p^t+M_p^t=c_p)
\end{multline}

\subsection{Traffic light control and optimal assignment of vehicles onto the lanes}
\label{sub-assign}

We are interested here in the equilibration of the two queue lengths with respect to the two parameters $\alpha$ and $\bar{r}:= r_N/r_M$.
We exclude here the case $r_N = r_M = 0$ where both lanes have green light, and where no queue is formed by assumption; and therefore, the
assignment onto the queues is meaningless.
We use notations $\mathbb E_{(\alpha,r_N)}(N^t):=\mathbb E(N^t)$ and $\mathbb E_{(\alpha,r_M)}(M^t):=\mathbb E(M^t)$ in order to emphasize the dependence of these two expectations on
the parameters $\alpha, r_N$ and $r_M$.
Let us now define $f(\alpha,\bar{r})$ as follows.

$$\begin{array}{ll}
 f(\alpha,\bar{r}) & := |\mathbb E_{(\alpha,r_N)}(N^t) - \mathbb E_{(\alpha,r_M)}(M^t)|/\mathbb E_{(\alpha,r_N)}(N^t) \\~~\\
                       & = |\mu_N - \mu_M|/\mu_N \\ ~~\\
                       & = \frac{\lambda|r_N(l_n+(1-\alpha)l_{nm}) - r_M(l_m+\alpha l_{nm})|}{\lambda r_N(l_n+(1-\alpha)l_{nm})} \\~~\\
                       & = \frac{|\bar{r}(l_n+(1-\alpha)l_{nm}) - (l_m+\alpha l_{nm})|}{\bar{r}(l_n+(1-\alpha)l_{nm})}.
 \end{array}
 $$

We are interested here in the minimization of $f(\alpha, \bar{r})$ with respect to the two parameters $\alpha$ and $\bar{r}$, which permits the equilibration of the two queue lengths.
Let us use the notations.
\begin{align}
  & r^*(\alpha) := \arg \min_{\bar{r}} f(\alpha, \bar{r}). \\
  & \alpha^*(\bar{r}) := \arg \min_{\alpha} f(\alpha, \bar{r}).
\end{align}

Proposition~\ref{prop:assignment} and Proposition~\ref{prop:assign2} below determines $r^*(\alpha)$ and $\alpha^*(\bar{r})$ respectively.

\begin{prop}
  $\forall \alpha \in [0,1], r^*(\alpha) = \frac{l_m+\alpha l_{nm}}{l_n+(1-\alpha)l_{nm}}$, and $f(\alpha, r^*(\alpha)) = 0$.
\label{prop:assignment}
\end{prop} 
\proof
$\forall \alpha \in [0,1], \bar{r} = \bar{r}_0 = \frac{l_m+\alpha l_{nm}}{l_n+(1-\alpha)l_{nm}}$ implies
$$\begin{array}{ll}
 \mathbb E_{(\alpha,r_N)}(N^t) & = r_N(\lambda_n+(1-\alpha)\lambda_{nm}) \\
                      & = (r_{M}\bar{r}) (\lambda_n+(1-\alpha)\lambda_{nm}) \\
                      & = r_{M}(\bar{r}_0(\lambda_n+(1-\alpha)\lambda_{nm})) \\
                      & = r_M(\lambda_m+\alpha\lambda_{nm}) = \mathbb E_{(\alpha, r_M)}(M^t).
\end{array}$$
Therefore, $f(\alpha, \bar{r}_0) = 0$. Thus, $r^*(\alpha) = \bar{r}_0=\frac{l_m+\alpha l_{nm}}{l_n+(1-\alpha)l_{nm}}$.
\endproof

\begin{prop} 
\label{prop:assign2}
  $\forall \bar{r} \geq 0$,
  $$\alpha^*(\bar{r}) = \max\left( 0, \min \left( 1, \frac{\bar{r}l_n + \bar{r} l_{nm}-l_m}{l_{nm}(\bar{r} + 1)}\right) \right).$$
  Moreover, if $\bar{r} \in I := [\frac{l_m}{l_n+l_{nm}},\frac{l_m+l_{nm}}{l_n}]$, then
  $$\alpha^*(\bar{r}) = \frac{\bar{r}l_n + \bar{r} l_{nm}-l_m}{l_{nm}(\bar{r} + 1)}, \text{ and } f(\alpha^*(\bar{r}), \bar{r}) = 0.$$
\end{prop}
\proof
For any $\bar{r}\geq 0$, $\alpha^*(\bar{r})$ is simply the argument of the minimization of $f(\alpha, \bar{r})$ with respect to $\alpha$, projected into the interval $[0,1]$.
In the case where $\bar{r} \in I := [\frac{l_m}{l_n+l_{nm}},\frac{l_m+l_{nm}}{l_n}]$, 
we can easily check that the constraint
$\alpha^*(\bar{r}) \in [0,1]$ is not activated, and then we do not need to project into the interval $[0,1]$.
Moreover, in this case, $\alpha^*(\bar{r})$ cancels $f(\alpha, \bar{r})$.
\endproof

We notice here that the calculus of the optimal assignment proportion $\alpha^*(\bar{r})$ of the vehicles going straight onto the lanes 
is done in deterministic at the macroscopic level (proportion of vehicular flow).
The realization of the optimal assignment proportion $\alpha^*(\bar{r})$ is done randomly at the microscopic level, as it has already been mentioned in section II.A.d:
every vehicle going straight is randomly assigned to the the lanes M and N with probabilities $\alpha$ and $(1-\alpha)$ respectively,
assuming that the vehicles going straight that can choose both lanes will choose the shortest queue.
By equilibrating the two queues with $r^*(\alpha)$ or $\alpha^*(\bar{r})$, we can avoid spill-back onto the links of the network
and by that reduce the risk of congestion.
For example, the optimal $r^*(\alpha)$ given in Proposition~\ref{prop:assignment} can be taken into account as an additional constraint in the optimization problem 
of the traffic light split on every intersection, in such a way that the optimal traffic light setting will automatically balance the queue lengths on the incoming links 
of the intersections, which should help to avoid spill-back at the network level.
\section{Simulation results, examples and discussion}
\label{sec-results}
We present in this section the results of numerical simulations we conducted with Veins framework~\cite{sommer2011bidirectionally}
which combines the microscopic road traffic simulator SUMO~\cite{SUMO2012} with the communication simulator OMNET++~\cite{Varga01theomnet++}. 
The road network is one simple junction with links composed of two incoming lanes described in Fig.~\ref{fig:junction}.
The junction is controlled by a traffic light with a cycle duration of $90$ s. 
The traffic demand is coming from West towards East, North and South.
We vary the arrival rates and turn ratios depending on the scenarios, as we mentioned in TABLE~\ref{tab:demand}.
The messages we use to detect the location of the vehicles are the Basic Safety Messages (BSM) that are sent in broadcast by every probe vehicle.
Given the road network topology and the data provided by the probe vehicles, we can know on which incoming/outgoing link each probe vehicle is located.

\subsection{Primary parameters estimation}
\label{sub-results-param}

\begin{figure}[htbp]
  \begin{center}  
    \includegraphics[width=0.7\linewidth,keepaspectratio]{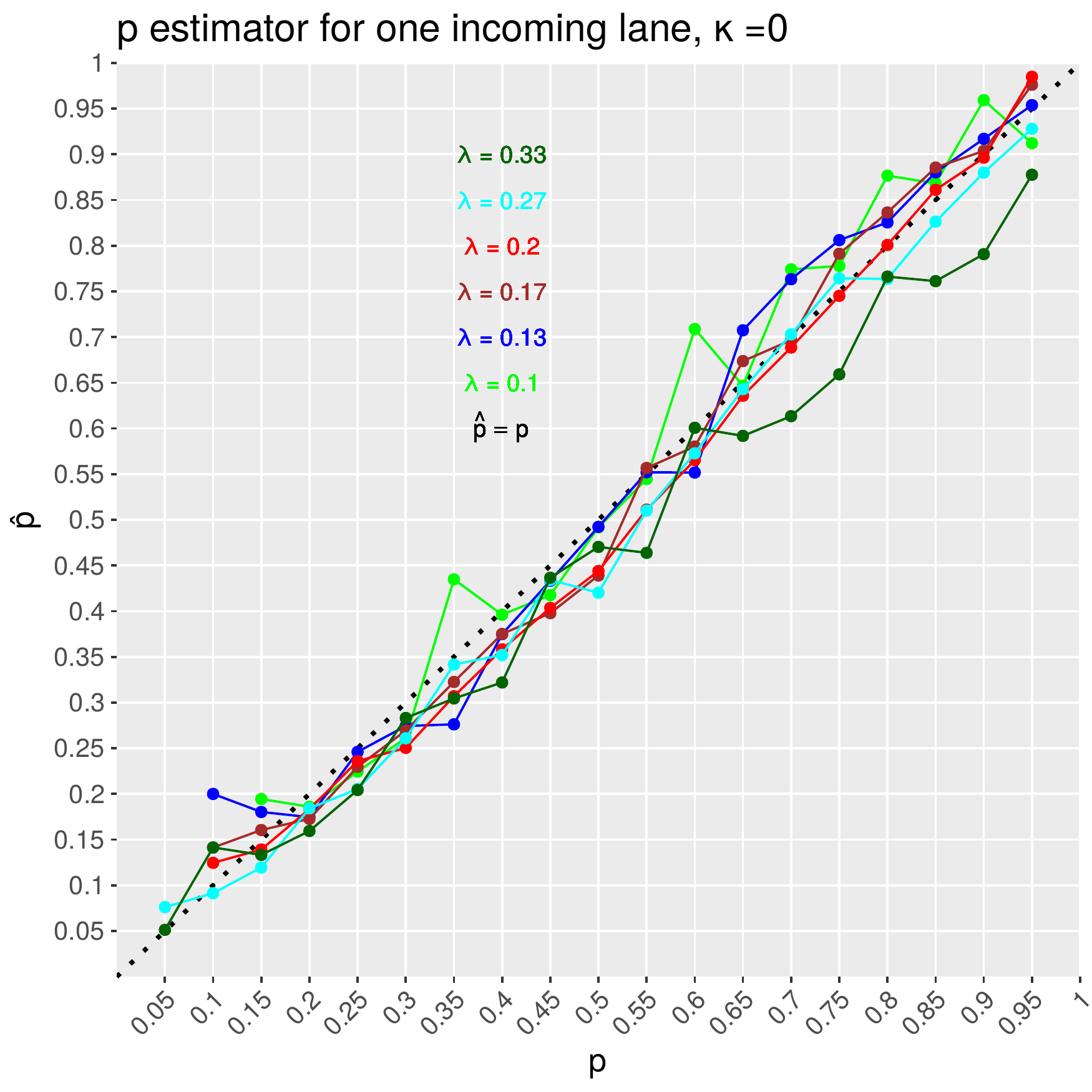}  
    \caption{Estimated penetration ratio $p$ for a one lane incoming link, depending on penetration ratio for various demand scenarios. 
    The arrival demand levels ($\lambda$) are given in (vehicles/s). Simulated time = 40 min.}
    \label{fig:p_estimated_one}
  \end{center}
\end{figure}

\begin{figure}[htbp]
  \begin{center}  
    \includegraphics[width=0.7\linewidth,keepaspectratio]{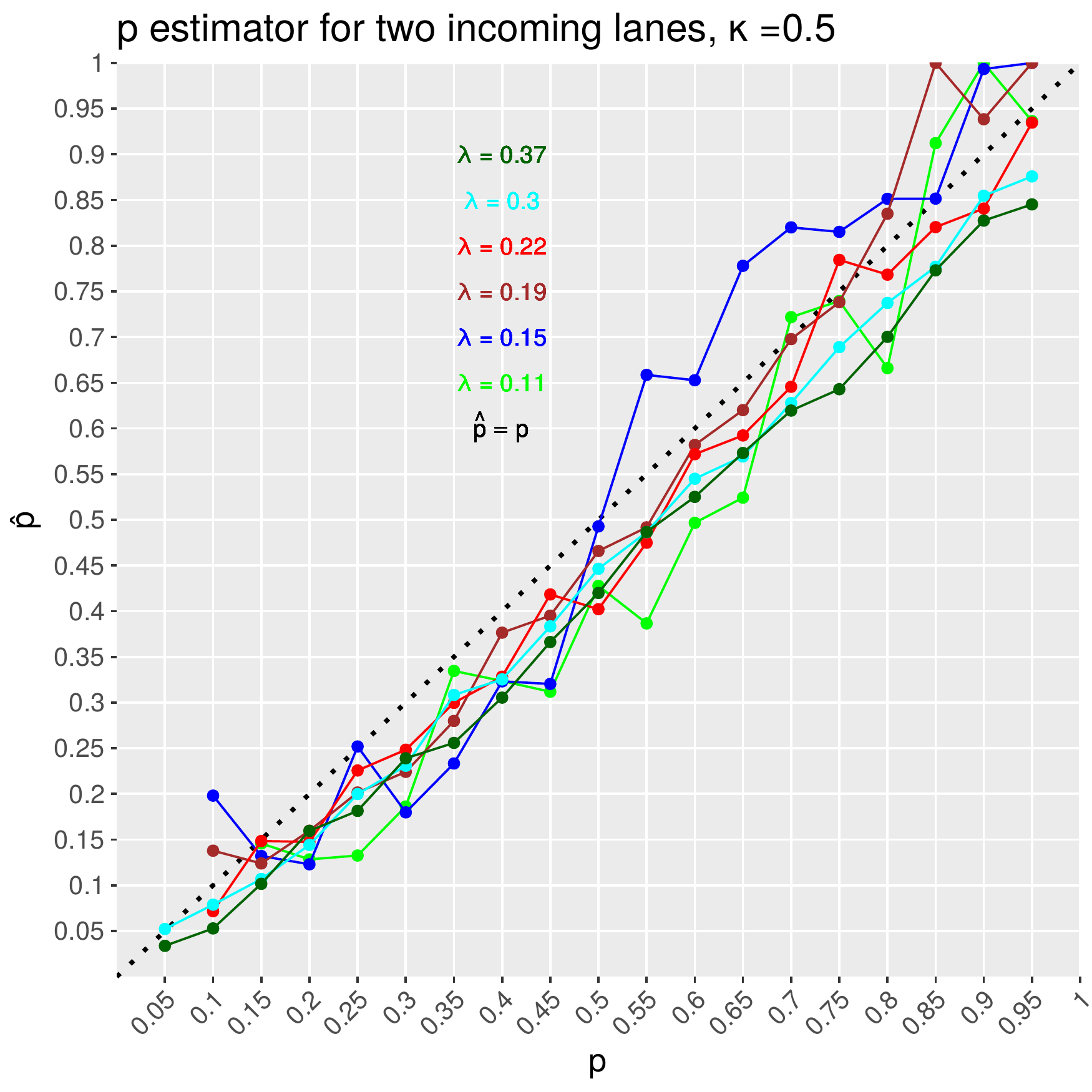}  
    \caption{Estimated penetration ratio $p$ for a two lanes incoming link, depending on penetration ratio for various demand scenarios. 
    The arrival demand levels ($\lambda$) are given in (vehicles/s). Simulated time = 40 min. $\kappa=0.5$.}
    \label{fig:p_estimated}
  \end{center}
\end{figure}
In this part, we illustrate estimation of primary parameters we proposed in section~\ref{sub-primary}.
Fig.~\ref{fig:p_estimated_one} represents the estimated penetration ratio $\hat{p}$, given by formula~(\ref{eq:p}),
associated in this figure to the real penetration ratio $p$, in the case of an incoming link of one lane.
Fig.~\ref{fig:p_estimated} represents the estimated penetration ratio $\hat{p}$,
given by formula~(\ref{eq:hatp2}), associated in this figure to the real penetration ratio $p$, in the case of an incoming link of two lanes.
Ideally, $\hat p=p$, forming a line of slope $1$ drawn in discontinuous black on the figure.

\begin{figure}[htbp]
  \begin{center}  
    \includegraphics[width=0.7\linewidth,keepaspectratio]{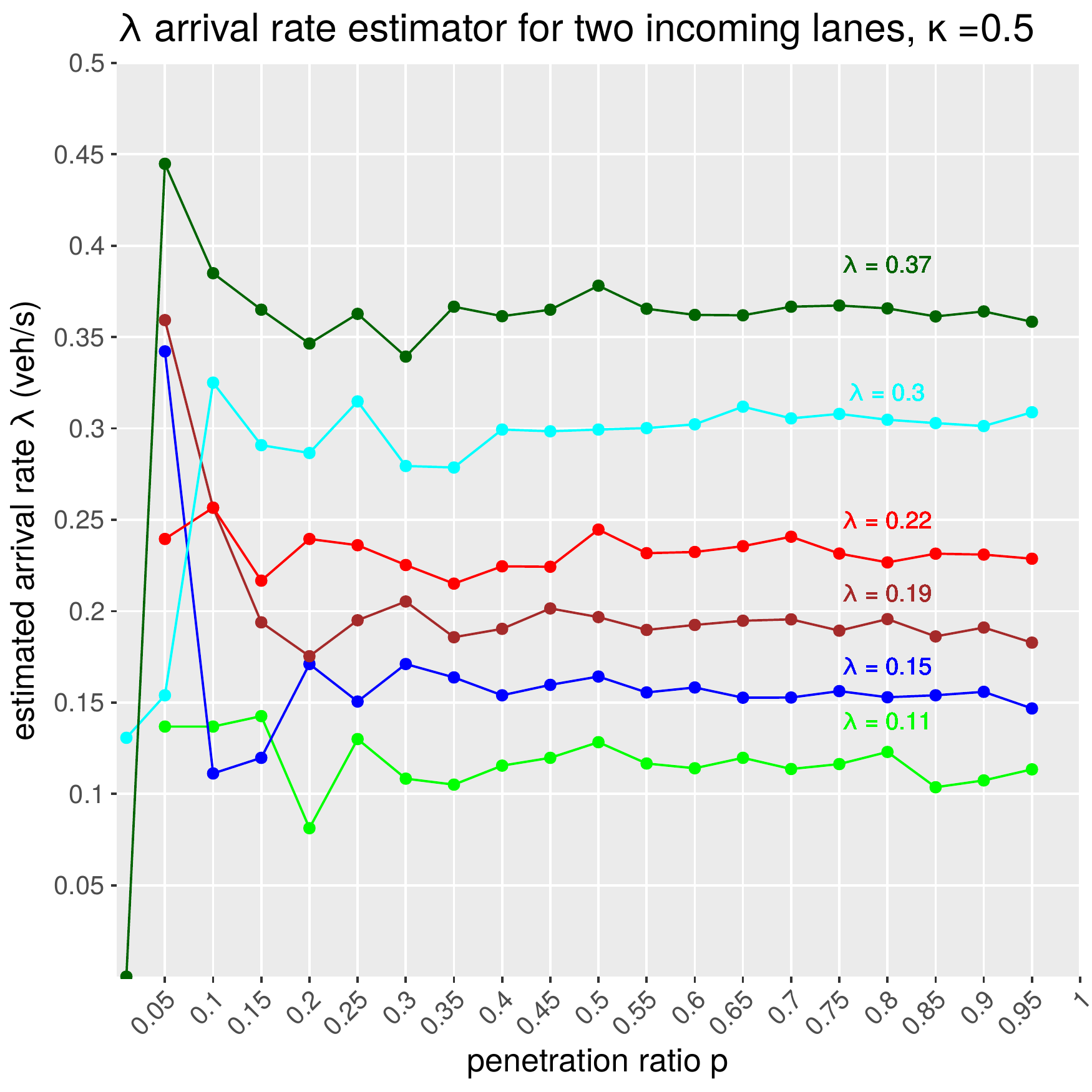}  
    \caption{$(\lambda_n+\lambda_m+\lambda_{nm})$ in vehicles/second for a two lanes incoming link, depending on penetration ratio for various demand scenarios. Simulated time = 40 min. $\kappa=0.5$.}
    \label{fig:lambda_estimated}
  \end{center}
\end{figure}
Fig.~\ref{fig:lambda_estimated} represents arrival rate estimated as given by formula~(\ref{eq:lambda2}) of section~\ref{sub-primary}
in the case of an incoming link of two lanes for $\kappa=0.5$.
We can see here that the estimation is better when $p$ gets higher.
As there are more data and as the arrival rate is higher, the estimation of arrival rate is more accurate.

\subsection{Probability distributions}
\label{sub-results-proba}
In this part, we illustrate the probability distribution queue lengths as proposed in section~\ref{sub-queue_2_lanes}.
We assume the demand is coming from the West of the junction as described in TABLE~\ref{tab:demand} with $r_N=r_M$ i.e. $\bar{r}=1$.

These different scenarios include different possibilities concerning the demand such as : symmetric (Scenario S3) or asymmetric arrivals (scenarios S1, S2, S4, S5). 
We vary the arrival demand $\lambda_n, \lambda_m$ and $\lambda_{nm}$ and derive the optimal $\alpha^*(1)$ by Proposition~\ref{prop:assign2}.
We obtain the values $\alpha^*(1)=[0.1,0.25,0.5,0.75,0.9]$.
We do not consider the cases where $\bar{r}\neq1$.
\begin{table}[htbp]
\centering
\begin{tabular}{|c|c|c|c|c|c|}
\hline
\textbf{Scenario} & S1 & S2 & S3 & S4 & S5 \\
\hline
$\alpha^*$  & 0.1 & 0.25 & 0.5 & 0.75 & 0.9 \\
\hline
$\lambda_{nm}$ & 125 & 100 & 50 & 100 & 125 \\
$\lambda_{m}$  & 200 & 125 & 200 & 75 & 100\\
$\lambda_{n}$  & 100 & 75 & 200 & 125 & 200\\
\hline
\textbf{Arrival rates} & \multicolumn{5}{|c|}{\textbf{Amount of vehicles for 1200 s}}\\
\hline
\end{tabular}
\caption{Demand for different scenarios (simulated time=1200 s) and for $r_N=r_M$}
\label{tab:demand}
\end{table}

\paragraph{Example: probability distribution of the two queues lanes without and with the information provided by the probe vehicles}
\begin{figure}[htbp]
  \begin{center}  
    \includegraphics[width=0.7\linewidth, keepaspectratio]{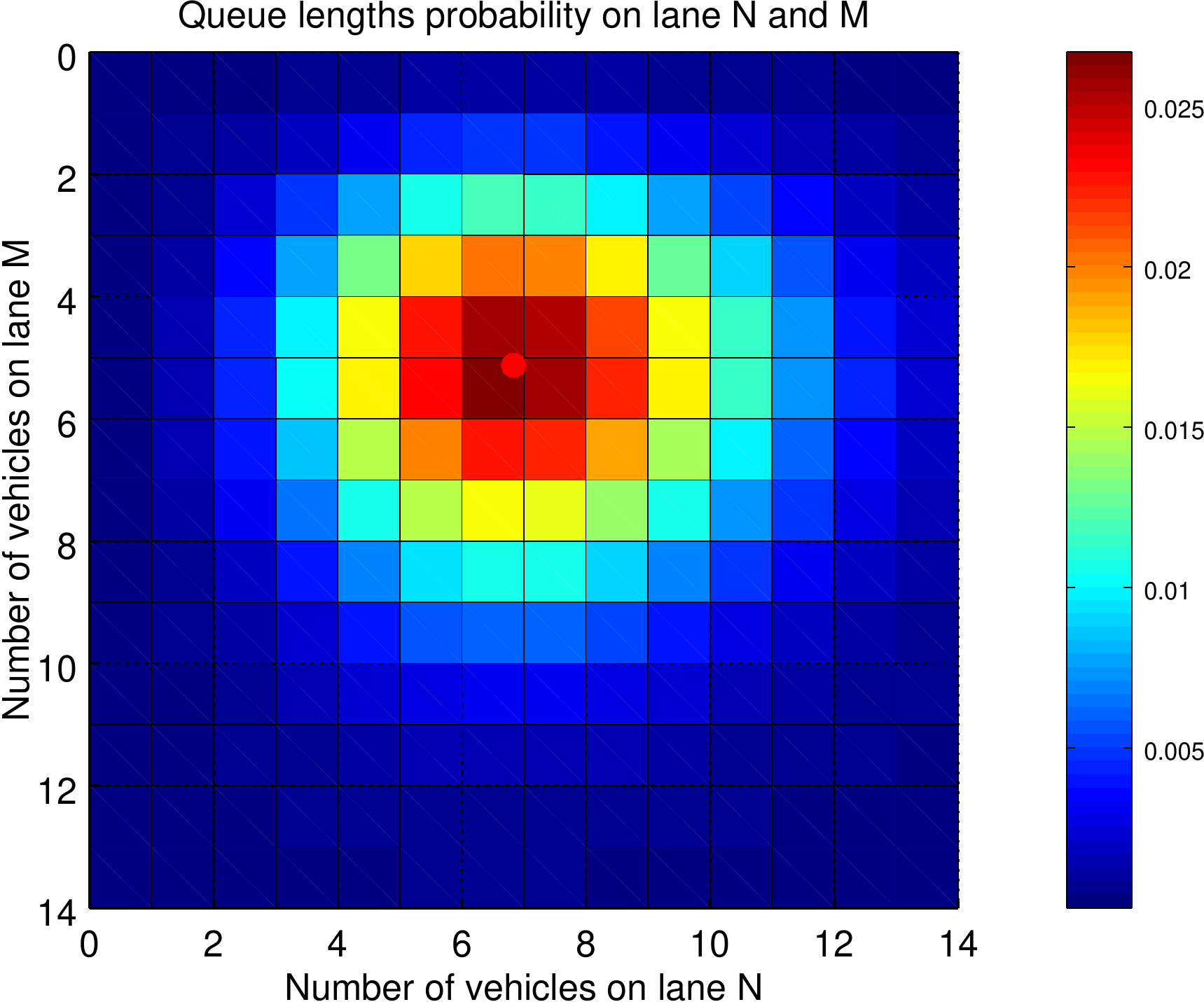}  
    
    \vspace{0.2cm}
    \includegraphics[width=0.7\linewidth, keepaspectratio]{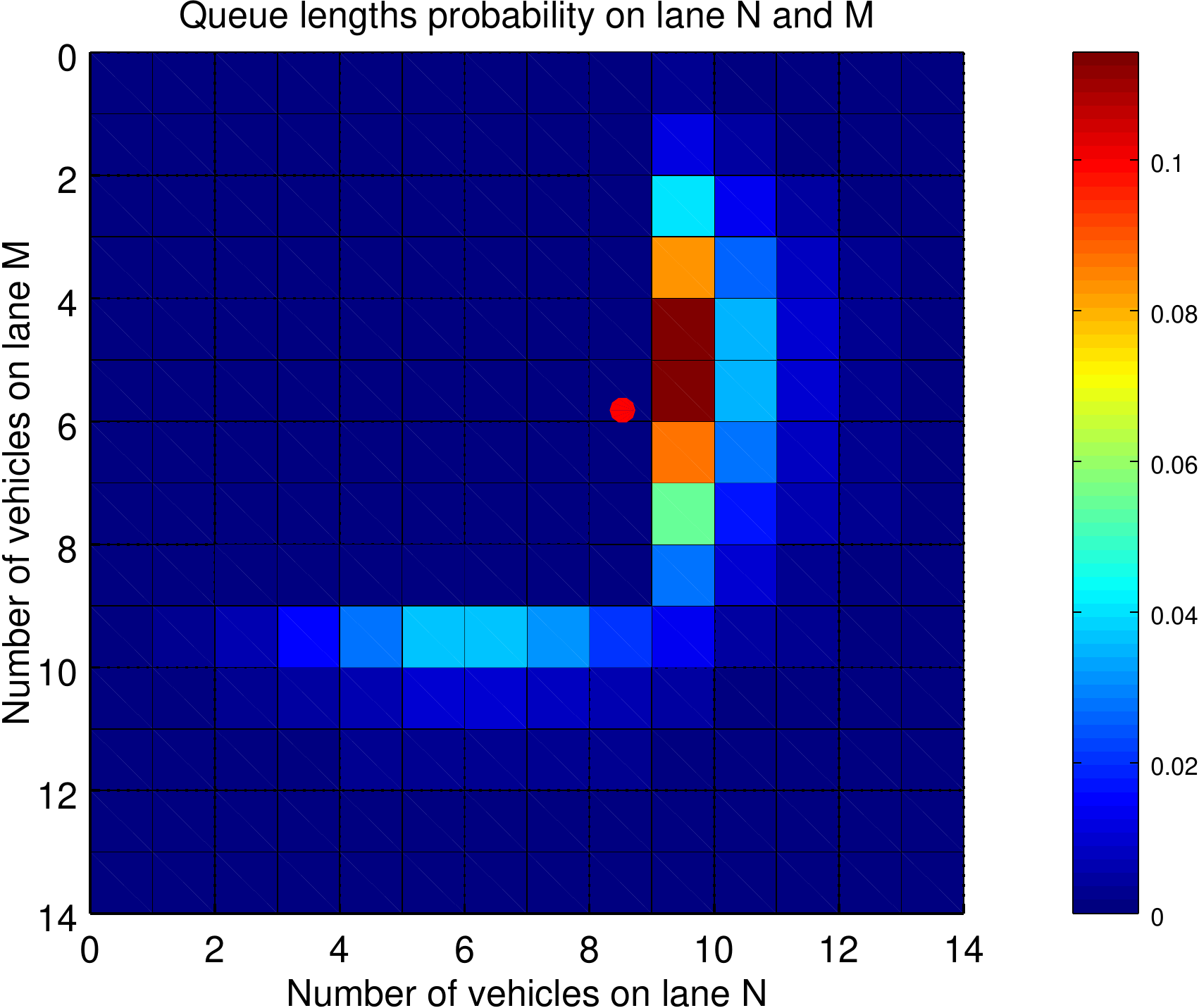}
    \caption{ On the top : probability law of Proposition~\ref{prop1}. On the bottom : probability law of Proposition~\ref{prop-2}.
	      Scenario with $\lambda_n=1/6, \lambda_m=1/12, \lambda_{nm}=1/24$ (veh/s)(thus $\alpha^*(1)=1$), $r(t)=41$ s, $p=0.55$, $c_p=8$, $l_p=9$, at time $t=760 s$.
	      Red dot is the expectation of the probability distribution. Simulation time=1200s.}
  \label{fig:matrixS5}	      
  \end{center}
\end{figure}
We draw on top of Fig.~\ref{fig:matrixS5} the probability distribution law $P(N^t,M^t)$ (Proposition~\ref{prop1}), 
and on bottom of Fig.~\ref{fig:matrixS5} the conditional probability distribution law $P(N^t,M^t | L_p, N_p+M_p)$ (Proposition~\ref{prop-2} ) for scenario of Fig.~\ref{fig:matrixS5}.
On top of Fig.~\ref{fig:matrixS5}, we can see that the total number of vehicles in the queue is estimated to $(N^t=6,M^t=4)$,
for a total of $10$ vehicles in the queue. There is an asymmetry in the distribution probability law because of the asymmetric demand and because the common flow 
is not strong enough to equilibrate the two queue lanes.
We can see on bottom of Fig.~\ref{fig:matrixS5} that the conditional distribution clearly discriminates the two queue lengths and keeps track of the asymmetry.
In this case, the parameters are $p=0.55$, $c_p=8$, $l_p=9$. We compute $\kappa=0.75$. Following the same ideas as above, we have $\hat{p}=(8/1.75-1)/(9-1)=0.45$.
It is probable that there are not many cars behind $l_p$, maybe $1$ vehicle.
Therefore, as $l_p=9$, the biggest lane should contain around $10$ vehicles.
Given the asymmetry of the distribution law $P(N^t,M^t)$,
the conditional probability $P(N^t,M^t | L_p, N_p+M_p)$ will favour the lane with the highest arrival rate (lane $N$).
Then the queue on lane $M$ should contain very few vehicles and will be around the same lane length estimation as in the top figure.
In this example, the conditional distribution probability calculus emphasizes the asymmetry of the two lanes.
\paragraph{Results for the scenarios of TABLE~\ref{tab:demand}}
For each scenario we measure the maximum and average queue lengths as estimated by SUMO microscopic road traffic simulator.
We have used different seeds for the random number generator of the simulators, taking the average value for TABLE~\ref{tab:resultsN}~and~\ref{tab:resultsM}.
We notice that SUMO queue length is measured in such
a way that any vehicle with a speed greater than $0.1$ m/s is not considered in the queue.
We vary $p\in[0,1]$ for each scenario and compute :
\begin{itemize}
 \item MAE(P2):=the mean absolute error between the estimated queue lengths as given by the estimator based on Proposition~\ref{prop-2} 
 and SUMO queue lengths on a subset of the data.
 \item MAE(P1):=the mean absolute error between the estimated queue lengths as given by $\hat{N}=\mu_n$, $\hat{M}=\mu_m$ (Proposition~\ref{prop1})
 and SUMO queue lengths on a subset of the data. 
 \item MAE($l_p$):=the mean absolute error between the estimated queue lengths as given by $\max(\hat{N},\hat{M})=l_p$ and $\min(\hat{N},\hat{M})=\kappa l_p$,
  for the queue lengths N and M, and SUMO queue lengths on a subset of the data.
\end{itemize}

\begin{table}[htbp]
\begin{tabular}{|c|c|c|c|c|c|c|}
\hline
\diagbox{Results}{Scenarios} & S1 & S2 & S3 & S4 & S5\\
\hline
Average SUMO&3.54 &3.21&4.33&3.21&4.77\\
Queue Length&     &    &    &      &	\\
\hline
Max SUMO    &10.07& 10.07&14.12&9.71&15.95\\
Queue Length&     &     &      &      &	\\
\hline
$p=0.05$ MAE(P2)&1.59		  & 1.11	 &1.38	   &1.19	   &1.84	 \\
MAE(P1)         &1.40		  & 1.14	 &1.45	   &1.20	   &1.97	 \\ 
MAE($l_p$)      &3.91		  & 3.71	 &4.38	   &3.71	   &5.61	 \\
MAPE(P2,R)(\%)  &34.00            &29.65         &25.89    &29.72          &29.28        \\   
\hline           	      	      	        	        	
$p=0.10$ MAE(P2)&1.51		  & 1.02	 &1.29	   &1.12	   &1.75	 \\
MAE(P1)         &1.40		  & 1.14	 &1.45	   &1.20	   &1.97	 \\ 
MAE($l_p$)      &2.01		  & 2.28	 &1.96	   &2.13	   &2.60	 \\ 
MAPE(P2,R)(\%)  &31.64            &26.50         &23.35    &26.05          &26.45        \\    
\hline           	      	      	        	        	
$p=0.15$ MAE(P2)&1.52		  & 1.04	 &1.28	   &1.06	   &1.58	 \\
MAE(P1)         &1.40		  & 1.14	 &1.45	   &1.20	   &1.97	 \\ 
MAE($l_p$)      &1.77		  & 1.52	 &1.64	   &1.53	   &1.95	 \\
MAPE(P2,R)(\%)  &29.33            &27.35         &20.35    &23.34          &20.71        \\    
\hline           	      	      	        	        	
$p=0.20$ MAE(P2)&1.47		  & 0.96	 &1.20	   &0.98	   &1.43	 \\
MAE(P1)         &1.40		  & 1.14	 &1.45	   &1.20	   &1.97	 \\ 
MAE($l_p$)      &1.85		  & 1.28	 &1.40	   &1.16	   &1.44	 \\
MAPE(P2,R)(\%)  &24.71            &25.40         &19.52    &19.78          &20.63        \\    
\hline           	      	      	        	        	
$p=0.50$ MAE(P2)&1.47		  & 0.80	 &0.98	   &0.79	   &1.10	 \\
MAE(P1)         &1.40		  & 1.14	 &1.45	   &1.20	   &1.97	 \\ 
MAE($l_p$)      &2.33		  & 1.14	 &1.24	   &0.99	   &0.89	 \\
MAPE(P2,R)(\%)  &28.48            &21.30         &15.53    &14.09          &15.68        \\    
\hline           	      	      	        	        	
$p=0.70$ MAE(P2)&1.33		  & 0.73	 &0.93	   &0.70	   &1.13	 \\
MAE(P1)         &1.40		  & 1.14	 &1.45	   &1.20	   &1.97	 \\ 
MAE($l_p$)      &2.48		  & 1.25	 &1.30	   &0.97	   &0.90	 \\
MAPE(P2,R)(\%)  &24.24            &17.19         &15.68    &12.36          &14.74        \\    
\hline           	      	      	        	        	
$p=0.90$ MAE(P2)&1.10		  & 0.74	 &0.97	   &0.70	   &1.21	 \\
MAE(P1)         &1.40		  & 1.14	 &1.45	   &1.20	   &1.97	 \\ 
MAE($l_p$)      &2.62		  & 1.39	 &1.39	   &1.04	   &0.95	 \\
MAPE(P2,R)(\%)  &19.51            &16.53         &15.00    &12.93          &17.30        \\   
\hline                                  
\end{tabular}
\caption{Results in the unit ''number of vehicles`` for queue on lane N, 
estimated vs SUMO queue length.}
\label{tab:resultsN}
\end{table}

\begin{table}[htbp]
\begin{tabular}{|c|c|c|c|c|c|c|}
\hline
\diagbox{Results}{Scenarios} & S1 & S2 & S3 & S4 & S5\\
\hline
Average SUMO&4.79&3.39&4.20&2.89&3.84\\
Queue Length&    &      &      &      &	\\                                                    
\hline
Max SUMO    &15.22&9.71&14.11&8.61&12.28\\
Queue Length&     &      &      &      &	\\
\hline
$p=0.05$ MAE(P2)&1.98	  &1.24     &1.39         &1.26         &1.36    \\
MAE(P1)         &2.02	  &1.25	    &1.35	  & 1.22	&1.30    \\     
MAE($l_p$)      &5.71	  &3.94	    &4.27	  & 3.23	&4.22    \\
MAPE(P2,R)(\%)  &21.08    &28.26    &22.48        &32.15        &29.79   \\    
\hline           	        	        	       	     	       
$p=0.10$ MAE(P2)&1.89	  &1.17	    &1.30	  & 1.25	&1.37	 \\
MAE(P1)         &2.02	  &1.25	    &1.37	  & 1.22	&1.30	 \\  
MAE($l_p$)      &2.81	  &2.53	    &1.93	  & 1.88	&1.93	 \\  
MAPE(P2,R)(\%)  &17.38    &24.20    &19.45        &30.06        &27.66   \\   
\hline           	        	        	       	           
$p=0.15$ MAE(P2)&1.68     &1.09	    &1.22	  & 1.20	&1.29	 \\
MAE(P1)         &2.02     &1.25	    &1.37	  & 1.22	&1.30	 \\  
MAE($l_p$)      &1.94     &1.54	    &1.45	  & 1.44	&1.68	 \\
MAPE(P2,R) (\%) &15.85    &19.08    &18.25        &27.27        &26.71   \\   
\hline           	        	        	       	     	       
$p=0.20$ MAE(P2)&1.55     &1.05	    &1.19	  & 1.12	&1.28	 \\
MAE(P1)         &2.02     &1.25	    &1.37	  & 1.22	&1.30	 \\  
MAE($l_p$)      &1.54     &1.25	    &1.33	  & 1.20	&1.58	 \\
MAPE(P2,R)(\%)  &17.17    &21.62    &17.82        &24.47        &31.12   \\   
\hline           	        	        	       	     	       
$p=0.50$ MAE(P2)&1.14     &0.90	    &1.03	  & 0.89	&1.23	 \\
MAE(P1)         &2.02     &1.25	    &1.37	  & 1.22	&1.30	 \\  
MAE($l_p$)      &0.89	  &1.15	    &1.41	  & 1.30	&2.00	 \\
MAPE(P2,R)(\%)  &11.89    &16.53    &16.11        &18.88        &27.11   \\   
\hline           	        	        	       	     	      
$p=0.70$ MAE(P2)&1.23	  &0.82	    &0.96	  & 0.77	&1.14	 \\
MAE(P1)         &2.02	  &1.25	    &1.37	  & 1.22	&1.30	 \\  
MAE($l_p$)      &0.96	  &1.13	    &1.44	  & 1.40	&2.16	 \\
MAPE(P2,R)(\%)  &13.90    &15.71    &14.25        &13.47        &25.14   \\   
\hline           	        	        	       	     	       
$p=0.90$ MAE(P2)&1.28	  &0.80	    &0.91	  & 0.72	&1.00	 \\
MAE(P1)         &2.02	  &1.25	    &1.37	  & 1.22	&1.30	 \\  
MAE($l_p$)      &0.96	  &1.19	    &1.56	  & 1.50	&2.30	 \\
MAPE(P2,R)(\%)  &16.89    &15.19    &13.81        &12.96        &21.44   \\   
\hline                                      
\end{tabular}
\caption{Results in the unit ''number of vehicles`` for queue on lane M, 
estimated vs SUMO queue length.
}
\label{tab:resultsM}
\end{table}

We give here the results for the scenarios of TABLE~\ref{tab:demand} in TABLE~\ref{tab:resultsN}~and~TABLE~\ref{tab:resultsM}.
We comment and emphasize some tendencies on the results.
We notice on these tables that the error is decreasing as $p$ tends to $1$ :
Proposition~\ref{prop-2} is getting more accurate as more data is given in input.
Even if the estimator based on $l_p$ (MAE($l_p$)) is getting more accurate as $p$ increases,
we notice that the performances of our estimator based on Proposition~\ref{prop-2} are generally better.
Furthermore, if we compare MAE(P1) and MAE(P2) 
we notice that the performances of Proposition~\ref{prop-2} based estimator are in general more accurate than the ones of the other estimators, 
especially when $p$ increases. 
We notice finally that MAE($l_p$) gives in general less accurate results than the estimators based on Propositions~\ref{prop1}~and~\ref{prop-2}.
\footnote{A significant source of error for this model is the fact that the estimation is done in real numbers, while the 
    measured number of vehicles is done in integer ones. In the general case, the error due to discretization is about 0.5 vehicles. This is very big in the cases
    where the queue length is small, in particular at the beginning of the red time. If for example, the queue length is 0.5 vehicles in average, then
    we have 100\% error, due only to the discretization.}.

We think that the main source of difference between our estimations and SUMO queue lengths is the assignment model.
Indeed, drivers do not always choose the shortest queue for their assignment in SUMO.
    In fact, in SUMO, there is a kind of thresholds on the difference between the queue lengths, beyond which drivers choose the shortest queue.
    Moreover, the drivers do not have the same behavior, 
    i.e. the choice is stochastic, i.e. the probability of taking the shortest queue increases with the threshold on the difference on the queue lengths.
    Another source of uncertainty and error is the variance of the arrival flows. Indeed, estimation of the queue length takes into account the 
    average arrival flow $\lambda$, but in the simulation there is a variance of the time arrivals, which is then retrieved as an error of measurement.
    
We also give for information the order of magnitude of the following indicator in TABLE~\ref{tab:resultsN}~and~TABLE~\ref{tab:resultsM}:
MAPE(P2,R):=the mean percentage absolute error between the estimated queue lengths as given by the estimator based on Proposition~\ref{prop-2}
 and SUMO queue lengths on a subset of the data, at the end of the red time.
 Indeed, taking into account the MAPE at the beginning of the red time is meaningless (since the queues are not formed yet).
This indicator rather emphasizes the difference between our assignment model $\alpha^*$ and SUMO assignment model.
This indicators varies between around 10\% to 30\% depending on the penetration ratio and the simulation scenarios.
\begin{figure}[htbp]
  \begin{center}  
    \includegraphics[width=\linewidth, keepaspectratio]{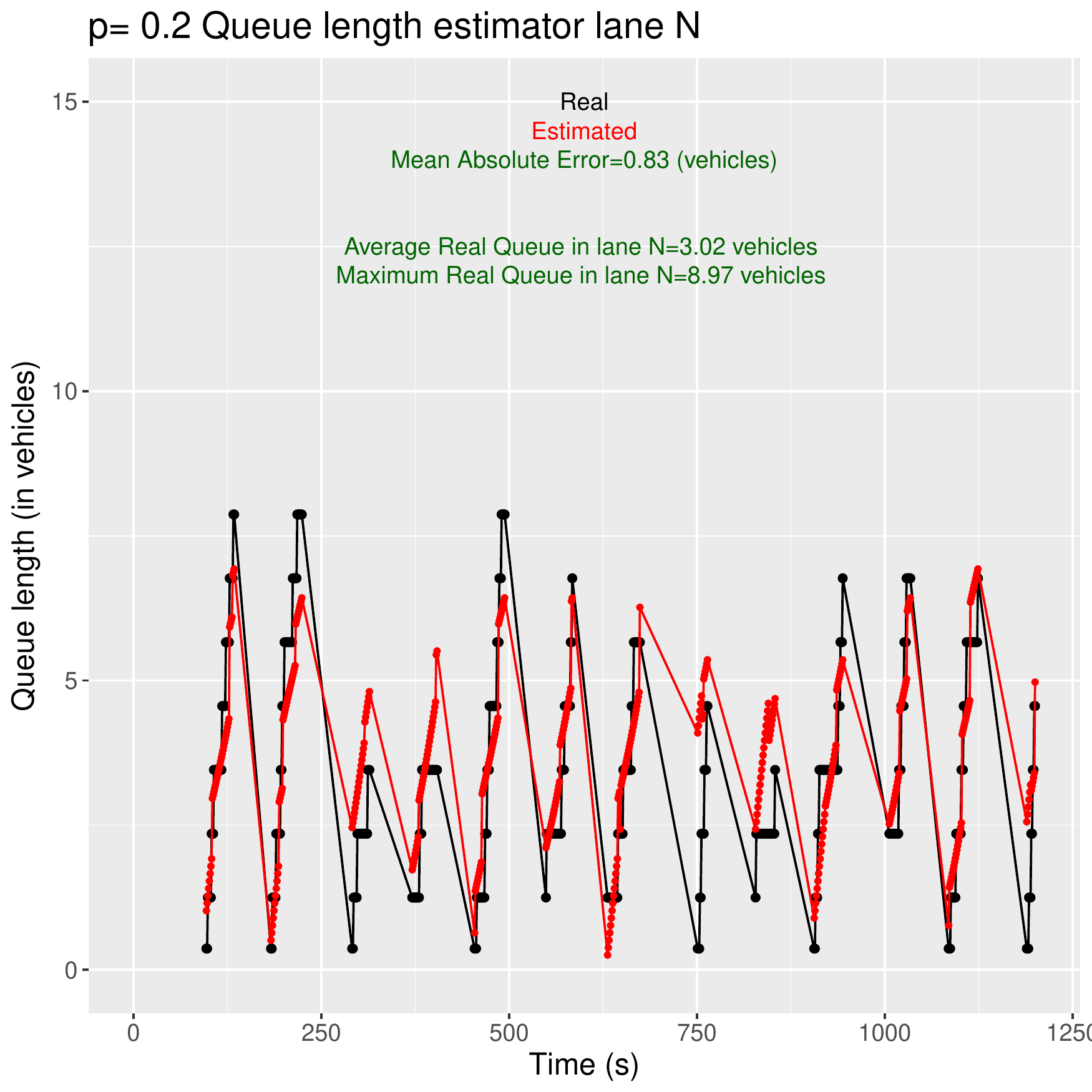}
    \includegraphics[width=\linewidth, keepaspectratio]{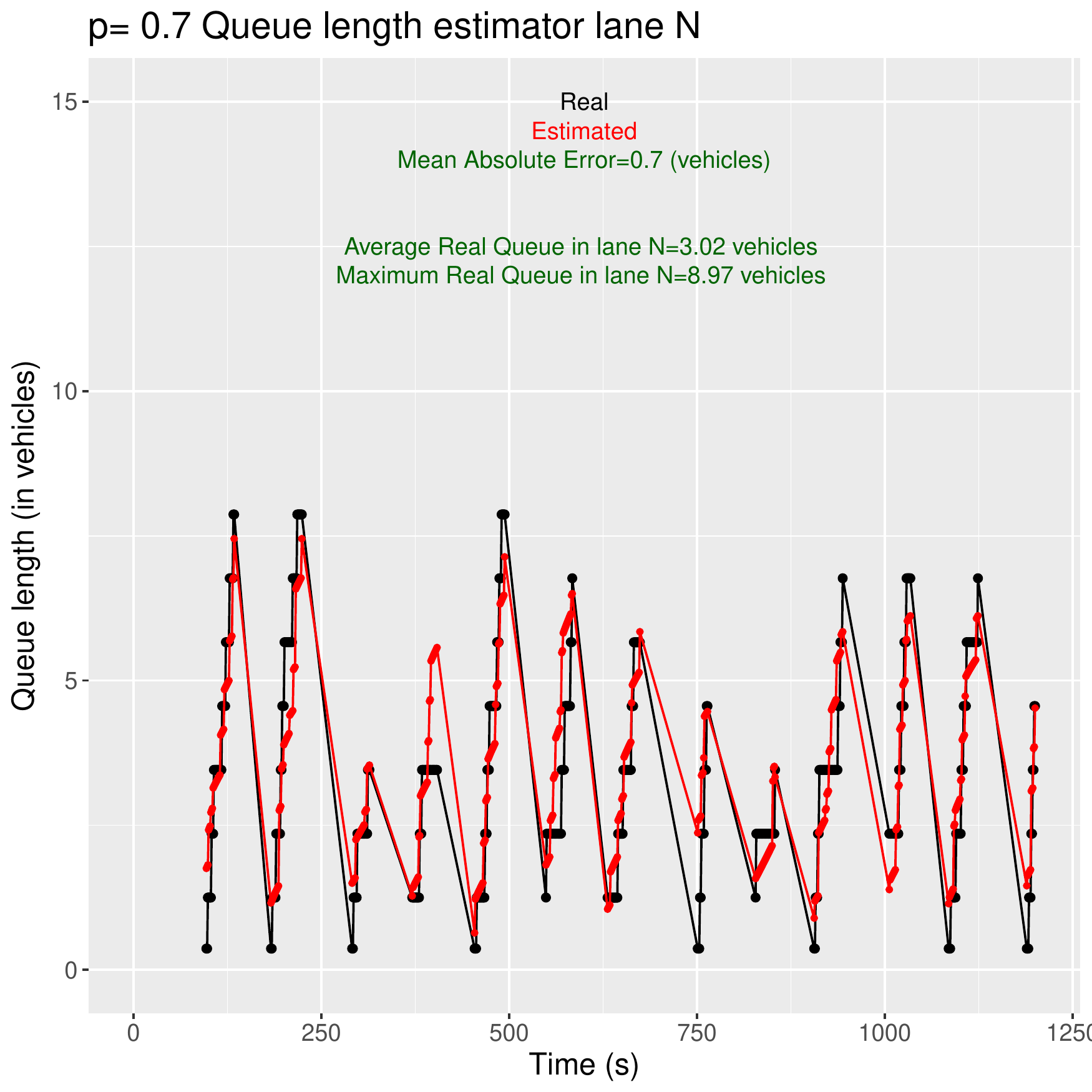}  
        \caption{Queue lengths estimator as given by Proposition~\ref{prop-2}, for varying $p=0.2$, $p=0.7$ and lane N, $\bar{r}=1$}
        \label{fig:simulationAssign}
  \end{center}
\end{figure}

\begin{figure}[htbp]
  \begin{center}  
    \includegraphics[width=\linewidth, keepaspectratio]{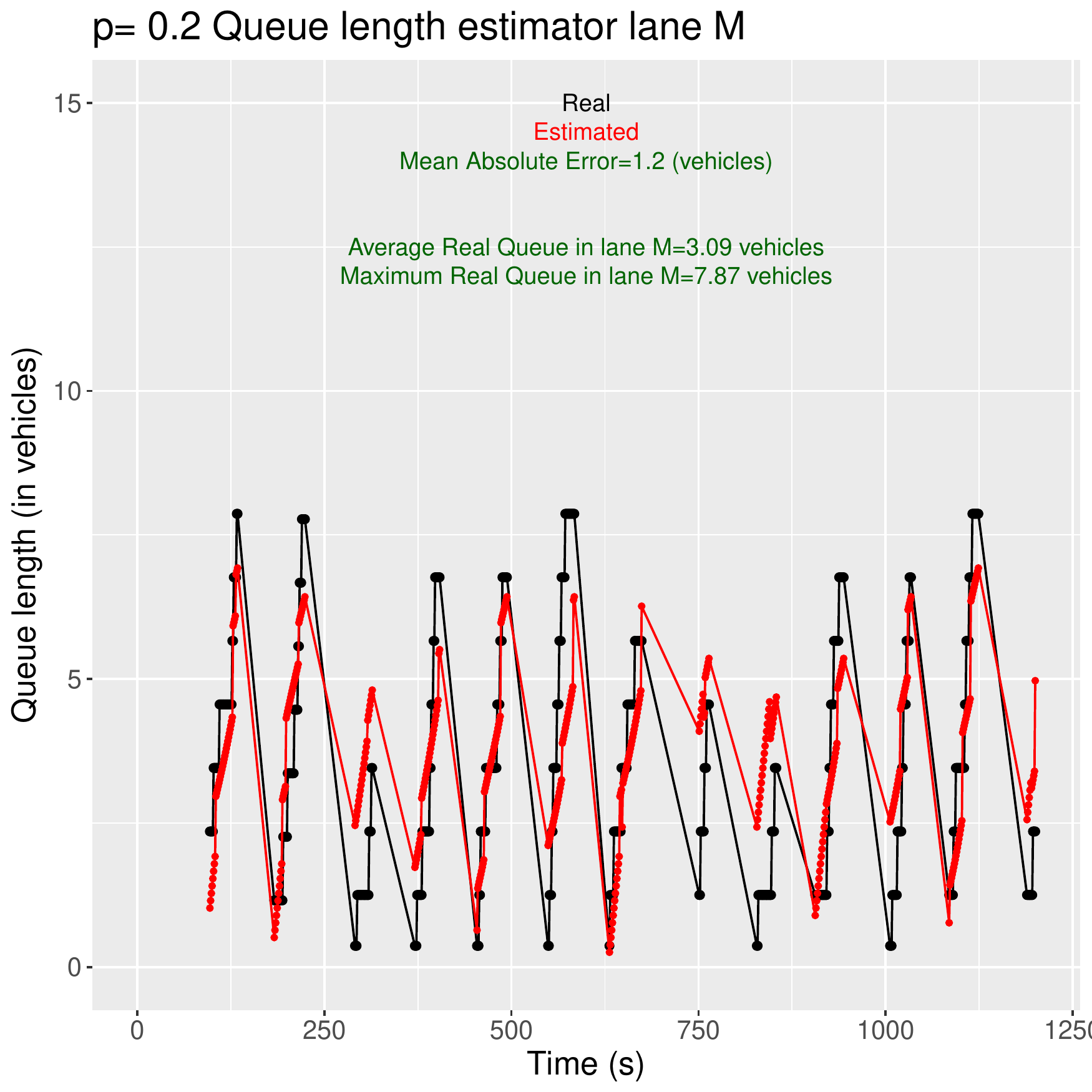} \hspace{5mm}
    \includegraphics[width=\linewidth, keepaspectratio]{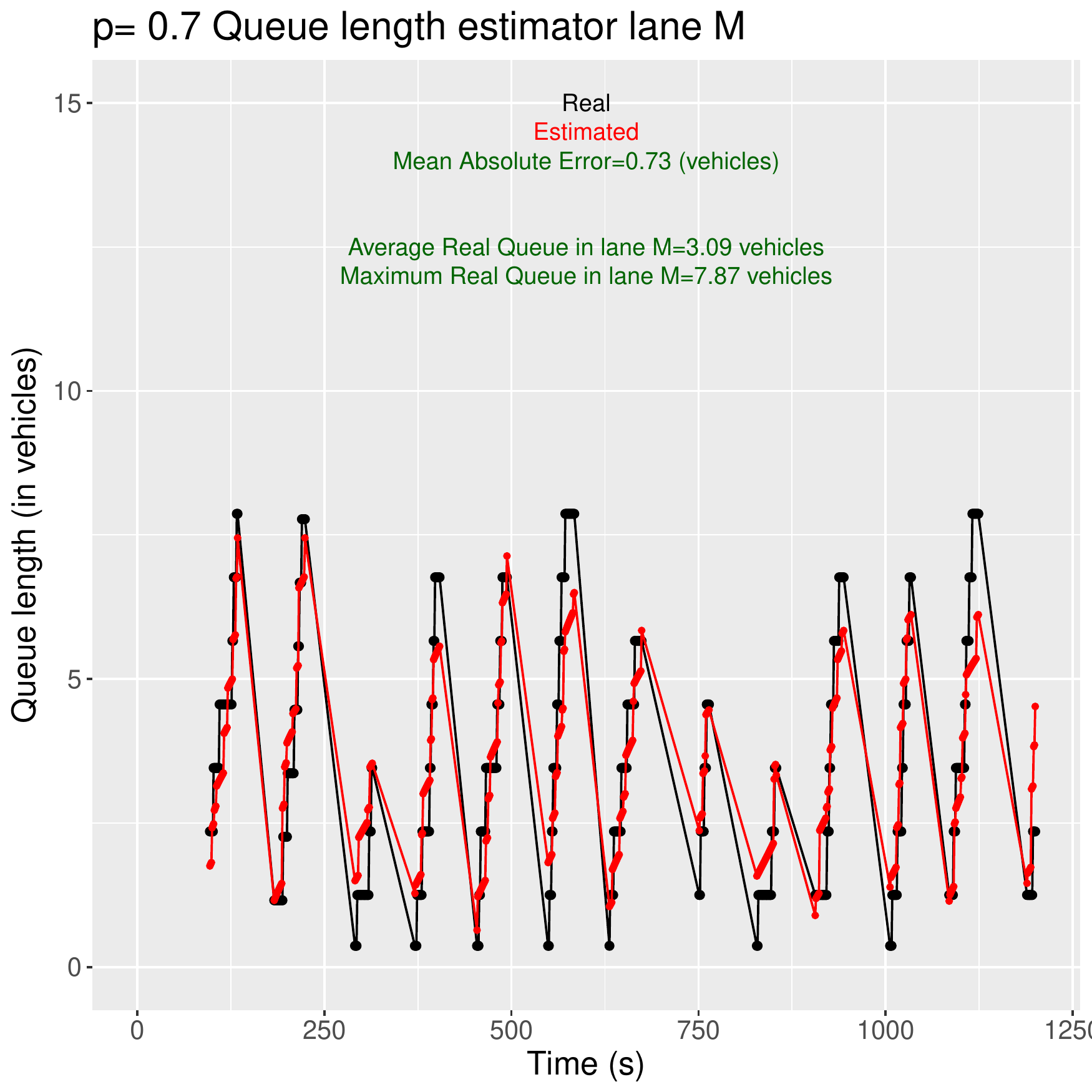}  
    \caption{Queue lengths estimator as given by Proposition~\ref{prop-2}, for varying $p=0.2$, $p=0.7$ and lane M, $\bar{r}=1$}
    \label{fig:simulationAssign2}
  \end{center}
\end{figure}

In Fig.~\ref{fig:simulationAssign}~and~Fig.~\ref{fig:simulationAssign2} we give the two lanes queue lengths in the scenario S4 where $r_N=r_M$.
We estimate queue lengths for $r_N>0$ and $r_M>0$.
We notice that the estimation is more accurate as $p$ gets higher.

\subsection{Traffic light control and vehicles assignment onto the lanes}
\label{sub-results-assign}
\begin{figure}[htbp]
  \begin{center}  
    \includegraphics[width=\linewidth,keepaspectratio]{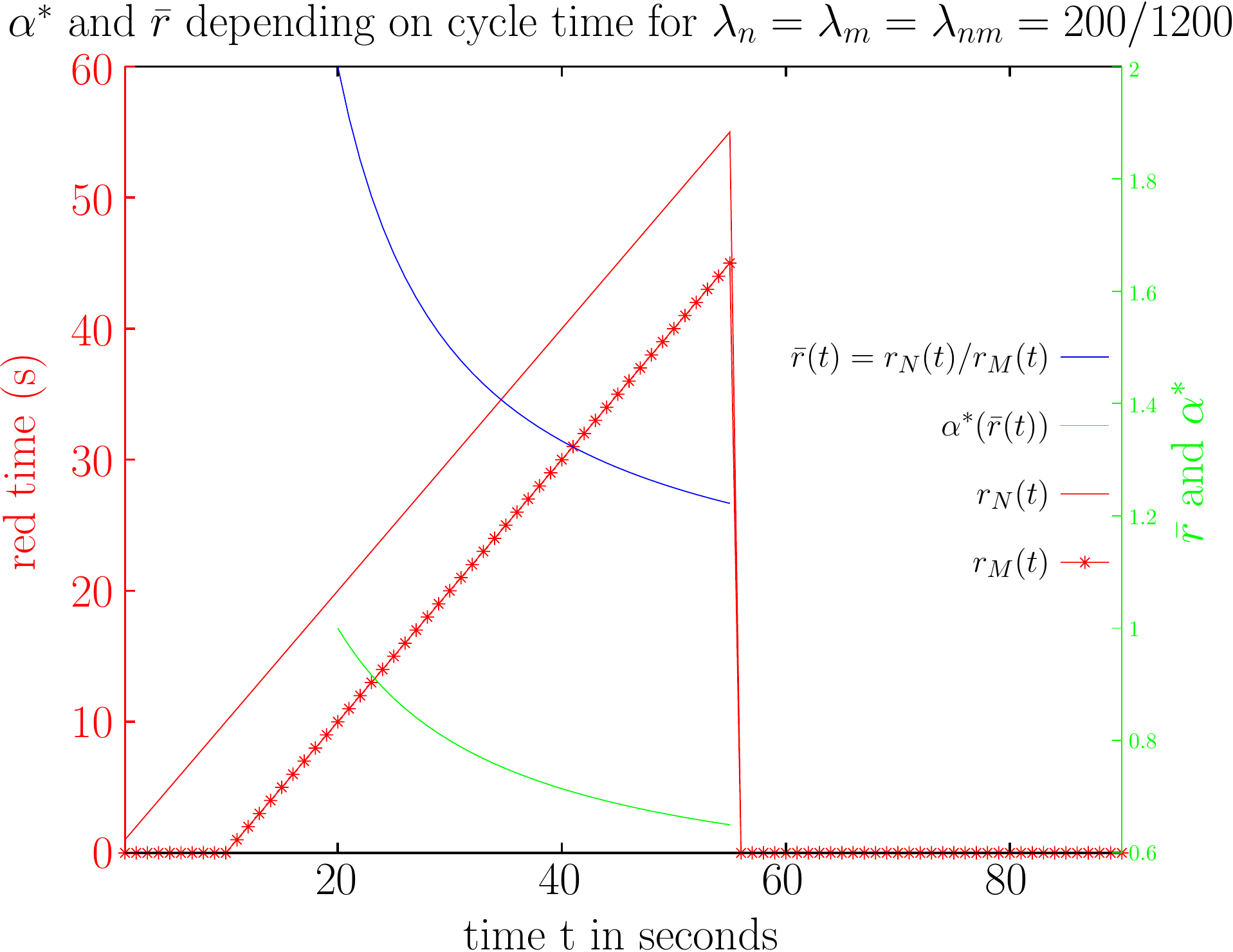}  
    \caption{$\alpha^{*}$ and $\bar{r}$ depending on cycle time of 90s for $\lambda_n=\lambda_m=\lambda_{nm}=\frac{200}{1200}=0.17$ vehicles/second.}
    \label{fig:rstar}
  \end{center}
\end{figure}

\begin{figure}[htbp]
  \begin{center}  
    \includegraphics[width=\linewidth,keepaspectratio]{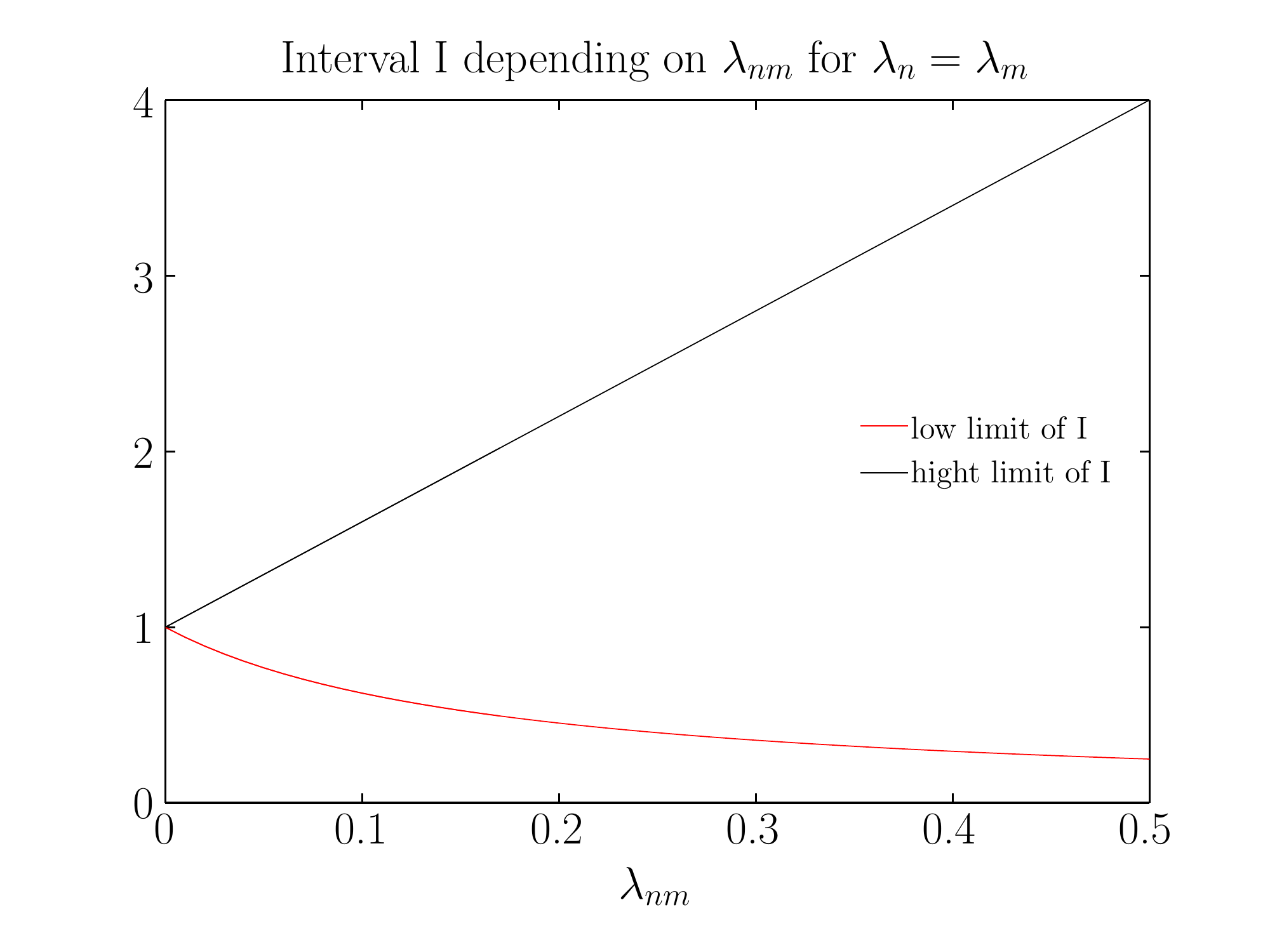}  
    \caption{$I=[\frac{\lambda_m}{\lambda_n+\lambda_{nm}},\frac{\lambda_m+\lambda_{nm}}{\lambda_n}]$ depending on $\lambda_{nm}$ for $\lambda_n=\lambda_m=1/6$}
    \label{fig:intervalI}
  \end{center}
\end{figure}
In this part, we assume symmetric demand, $\lambda_n=\lambda_m=\lambda_{nm}=0.17$ vehicles/second.
We assume the traffic light cycle includes a phase of 8 seconds, where green light is given to lane $M$, while red light is given to lane $N$.
Hence, the red duration on lanes $N$ and $M$ are different : $r_N\neq r_M$, $\bar{r}\neq 1$.
We also assume $\alpha=\alpha^*(\bar{r})$ given by Proposition~\ref{prop:assign2}.

On Fig.~\ref{fig:rstar} we draw $\bar{r}$ in blue and $\alpha^*(\bar{r})$ in green depending on time.
At the beginning of the cycle, the red is only for the lane $N$. The lane $M$ is at this time with green light.
Then, the two red lights are simply increasing as a line of slope~$1$.
Starting from 20 seconds, $\bar{r} \in I$ and $\alpha^*(\bar{r})=1$. This is because all the vehicles are assigned to lane $M$ 
which is the shortest queue (we recall it was at green light until then).
Then, $\alpha^*(\bar{r})$ decreases slowly to reach approximately $0.6$  which means that the two queue lengths are more equilibrated
as the red durations on lanes $N$ and $M$ are getting less different. We notice that $\alpha^*(\bar{r})$ would tend to $1/2$ if the red time goes to infinity.
Concerning $\bar{r}=r_n/r_m$, it is representing how the difference in the red lights durations is decreasing as time is going on.
$\bar{r}$ decrease is due to a constant 
offset (corresponding to the duration where lane $M$ is at green light while lane $N$ is at red light) which becomes less significant as the red light durations are increasing.
We notice that $\bar{r}$ would tend to $1$ if the red time goes to infinity.

The interval $I=[\frac{\lambda_m}{\lambda_n+\lambda_{nm}},\frac{\lambda_m+\lambda_{nm}}{\lambda_n}]$ as a function of $\lambda_{nm}$ is represented on Fig.~\ref{fig:intervalI}.
We notice on Fig.~\ref{fig:intervalI} that as the common flow $\lambda_{nm}$ gets higher, the interval I gets larger.
Therefore, as the common flow $\lambda_{nm}$ gets higher, there is more freedom to assign the vehicles onto a lane or another.

\subsection{Communication network performances}
\label{sub-results-comm}
We give here some information on the communication network performances that we have measured in simulation.
We have checked that the communication performances are not disturbing the estimation of the queue lengths.
As the order of magnitude of the end-to-end-delay
\footnote{The end-to-end delay is a communication indicator of performance
that measures the delay from the time a message is sent from a communicating vehicle until the time it is received by the receiver (in our case the receiver is the RSU).} 
is very low (around $0.2$ ms), we don't expect significant consequence on the queue length estimation application
as it could happen in scenarios where more vehicles would communicate, and cause significant delays such as described in \cite{NguyenVanPhu8005594}.
\FloatBarrier
\section{Conclusion and perspectives}
\label{sec-conclusion}
In this paper, we have proposed a method for the estimation of urban traffic state.
We give estimations for the penetration ratio of probe vehicles and for the vehicles arrival rate, on any link of the road network.
Knowing the arrival rate of the incoming flow and its composition, we have computed the joint probability distribution of the queue lengths in the case of two lanes link.
For this purpose, we have proposed a simple assignment model of vehicles onto the lanes.
In addition, we have refined the probability distribution of the queue lengths with the information provided by the probe vehicles.
A control of the traffic light has been proposed in order to balance the queues of the two lanes.
Moreover, we have proposed a formula for computing the optimal assignment of the vehicles onto the lanes.
Numerical simulations have been conducted with Veins framework, and the work presented here has been evaluated.
Road traffic control could benefit from the queue length estimations we presented in the present paper, in order to improve travel conditions.
We think the ideas we have given in this paper could be extended to a link of any number of lanes.
\bibliographystyle{IEEEtran}
\bibliography{./tits}

\appendices
\section{Calculus of the bias of $p$ estimator}
\label{Appendix:p}
We consider queue lengths on lanes $N$ and $M$ respectively equal to $n$ and $m$.
    We propose :
    \begin{equation}
    \hat{p}=\frac{c_p}{n+m}
      \label{eq:p_appendix}
      \end{equation}
      By the way, in our case, the length $n$ of queue $N$ can be estimated with the number of arrivals on lane $N$ during $r_N(t)$ which is $\mu_n$.
    As $\mu_n+\mu_m=\max(\mu_n,\mu_m)+\min(\mu_n,\mu_m)$ and by estimating $\max(\mu_n,\mu_m)=l_p+i_c/p$, where $0\leq i_c \leq 1$ 
    and $i_c/p$ represents the backlog of the queue (unequipped vehicles following $l_p$). 
    We want to determine $0\leq i_c \leq 1$ and compute $\hat{p}$ such that the estimation of $p$ is without bias.
    We can write :
    \begin{equation}
    \mu_n+\mu_m=\max(\mu_n,\mu_m)\left(1+\frac{\min(\mu_n,\mu_m)}{\max(\mu_n,\mu_m)}\right)
      \end{equation}
      \begin{equation}
    \mu_n+\mu_m=(l_p+i_c/p)(1+\kappa)
      \label{eq:mu_appendix}
      \end{equation}
      We introduce $c_\kappa=c_p/(1+\kappa)$ and replace~(\ref{eq:mu_appendix}) in~(\ref{eq:p_appendix}).
      Finally, we get the following equation : 
      \begin{equation}
	\hat{p}=\frac{c_\kappa}{l_p+i_c/p}
      \end{equation}
      \begin{equation}
	\hat{p}=(c_\kappa-i_c)/l_p
	\label{eq:hatp_appendix}
      \end{equation}
      We know from~\cite{COMERT2016502} how to compute the expectation of $\hat{p}$ and we follow the same ideas below :
      \begin{multline}
      P(L_p^t=l_p,N_p^t+M_p^t=c_p)=P(N_p^t+M_p^t=c_p|L_p^t=l_p)\\
      P(L_p^t=l_p)\\
      \end{multline}
      \begin{multline}
      P(L_p^t=l_p,N_p^t+M_p^t=c_p)=\binom{l_p-1+\min(l_p,n,m)}{c_p-1}\\
      p^{c_p-1}(1-p)^{l_p-1+\min(l_p,n,m)}P(L_p^t=l_p)\\
      \end{multline}
      \begin{multline}
      \mathbb E (\frac{c_p/(1+\kappa)-i_c}{l_p})=\sum_{l_p\geq 1}^{}\sum_{c_p=1}^{l_p-1+\min(l_p,n,m)}
      \frac{c_p/(1+\kappa)-i_c}{l_p}\\
      \binom{l_p-1+\min(l_p,n,m)}{c_p-1}\\
      p^{c_p-1}(1-p)^{l_p-1+\min(l_p,n,m)}P(L_p^t=l_p)\\
      \end{multline}
      \begin{multline}
      =\sum_{l_p\geq 1}^{}\frac{1}{l_p(1+\kappa)}\sum_{c_p=1}^{l_p-1+\min(l_p,n,m)}(c_p-1+1-i_c(1+\kappa))\\
      \binom{l_p-1+\min(l_p,n,m)}{c_p-1}\\
      p^{c_p-1}(1-p)^{l_p-1+\min(l_p,n,m)}P(L_p^t=l_p)\\
      \end{multline}
      To derive the next equation we use two arguments :
      \begin{itemize}
      \item the expectation of a binomial probability distribution law $\mathbb E(\mathbb B (n_x,p))=n_xp$ with $n_x=l_p-1+\min(l_p,n,m)$ in our case.
      \item and the formula of Newton $(a+b)^{m_x}=\sum_{k=0}^{m_x} \binom{m_x}{k}a^k b^{m_x-k}$, with $a=p$ and $b=1-p$ in our case.
      \end{itemize}  
      \begin{multline}
      \mathbb E (\frac{c_p/(1+\kappa)-i_c}{l_p})=\sum_{l_p\geq 1}^{}\frac{1}{l_p(1+\kappa)}\\
      \left(p(l_p-1+\min(l_p,n,m))+1-i_c(1+\kappa)\right)\\
      P(L_p^t=l_p)\\
      \end{multline}
      We replace $\min(l_p,n,m)=\kappa\max(l_p,n,m)=\kappa(l_p+i_c/p)$ :
      \begin{multline}
      \mathbb E (\frac{c_p/(1+\kappa)-i_c}{l_p})=\sum_{l_p\geq 1}^{}\frac{1}{l_p(1+\kappa)}(p(l_p-1+\kappa(l_p+i_c/p))+\\
      1-i_c(1+\kappa))\\
      P(L_p^t=l_p)\\
      \end{multline}
	\begin{multline}
      \mathbb E (\frac{c_p/(1+\kappa)-i_c}{l_p})=\frac{p}{1+\kappa}\mathbb E(\frac{l_p-1}{l_p}) + \frac{p\kappa}{1+\kappa} +\\
      \frac{i_c\kappa}{1+\kappa}\mathbb E(\frac{1}{l_p})+\frac{1-i_c(1+\kappa)}{1+\kappa}\mathbb E(\frac{1}{l_p})\\
      \end{multline}
      \begin{equation}
      \mathbb E (\frac{c_p/(1+\kappa)-i_c}{l_p})= p + \frac{\mathbb E(\frac{1}{l_p})}{1+\kappa}\left(1-p-i_c\right)  
      \end{equation}
      To get an estimator without bias, we write :
      \begin{equation}
      \mathbb E (\frac{c_p/(1+\kappa)-i_c}{l_p})= p  
      \end{equation}
      Solving this equation gives :
      \begin{equation}
      i_c=1-p
      \end{equation}
      Finally, we replace $i_c$ in~(\ref{eq:hatp_appendix}) :
      \begin{equation}
      \hat{p}=\frac{c_\kappa-(1-p)}{l_p}
      \end{equation}
      \begin{equation}
      \hat{p}=\frac{c_\kappa-1}{l_p-1}
      \end{equation}
\end{document}